\documentclass[12pt]{article}
\usepackage{axodraw}

\makeatletter

\renewcommand{\section}{\@startsection{section}{1}{\z@}%
                                    {-7ex \@plus -1ex \@minus -.2ex}%
                                    {2.5ex \@plus.2ex}%
                                    {\normalfont\large\scshape\centering}}                     \renewcommand{\subsection}{\@startsection{subsection}{2}{\z@}%
                                       {-5ex \@plus -1ex \@minus -.2ex}%
                                       {1.5ex \@plus.2ex}%
                            {\normalfont\normalsize\scshape}}
\newcommand{\sectionname}{}

\renewcommand\@seccntformat[1]{\ignorespaces\csname #1name\endcsname\space
                               \csname the#1\endcsname.\quad}   
\renewcommand{\appendix}{\par
  \setcounter{section}{0}%
  \setcounter{subsection}{0}%
  \renewcommand{\thesection}{\@Alph\c@section}%
  \renewcommand{\sectionname}{\appendixname}}
\newdimen\captionmargin 
\newdimen\captionindent 
\newdimen\captionwidth 
\newcommand{\captionfont}{\slshape}
\newcommand\@captionlabel[1]{\textsc{#1:}\space}
\long\def\@makecaption#1#2{%
  \vskip\abovecaptionskip  
  \captionwidth\hsize 
  \advance\captionwidth -2\captionmargin
  \sbox\@tempboxa{\@captionlabel{#1}\captionfont #2}%
  \ifdim \wd\@tempboxa >\captionwidth
    \ifdim\captionindent>\z@ 
      \advance\captionwidth -\captionindent
      \hskip\captionindent
    \fi
    \hskip\captionmargin
    \parbox[t]{\captionwidth}{\leavevmode\hskip-\captionindent
      \@captionlabel{#1}\captionfont #2}%
  \else
    \global \@minipagefalse
    \hb@xt@\hsize{\hfil\box\@tempboxa\hfil}%
  \fi
  \vskip\belowcaptionskip}
\def\eqnarray{%
   \stepcounter{equation}%
   \def\@currentlabel{\p@equation\theequation}%
   \global\@eqnswtrue
   \m@th
   \global\@eqcnt\z@
   \tabskip\@centering
   \let\\\@eqncr
   $$\everycr{}\halign to\displaywidth\bgroup
       \hskip\@centering$\displaystyle\tabskip\z@skip{##}$\@eqnsel
      &\global\@eqcnt\@ne$\;\hfil{##}$\hfil
      &\global\@eqcnt\tw@$\;\displaystyle{##}$\hfil\tabskip\@centering
      &\global\@eqcnt\thr@@ \hb@xt@\z@\bgroup\hss##\egroup
         \tabskip\z@skip
      \cr
}
\newcommand{\eqref}[1]{(\ref{#1})}
\newcommand{\Chref}[1]{\expandafter\MakeUppercase\chaptername~\ref{#1}}
\newcommand{\Secref}[1]{\expandafter\MakeUppercase\secrefname~\ref{#1}}
\newcommand{\Appref}[1]{\expandafter\MakeUppercase\appendixname~\ref{#1}}
\newcommand{\Figref}[1]{\expandafter\MakeUppercase\figurename~\ref{#1}}
\newcommand{\Tbref}[1]{\expandafter\MakeUppercase\tablename~\ref{#1}}
   \let\secref\Secref  \let\appref\Appref
\let\figref\Figref  \let\tbref\Tbref
\newcommand{\secrefname}{Section}
\DeclareSymbolFont{AMSb}{U}{msb}{m}{n}
\DeclareSymbolFontAlphabet{\mathbb}{AMSb}
\DeclareSymbolFont{boldletters}{OML}{cmm}{b}{it}
\newcommand{\inv}{^{\raise.15ex\hbox{${\scriptscriptstyle -}$}\kern-.05em 1}} 

\newcommand{\vev}[1]{\langle #1 \rangle}                  
\newcommand{\Vev}[1]{\Bigl\langle #1 \Bigr\rangle}
\newcommand{\ext}[1][]{\mathop{\raisebox{.2ex}{$\textstyle\bigwedge$}^{#1}}}
\newcommand{\del}{\partial}

\newcommand{\bdA}{\bar\partial_{\!A}}

\newcommand{\e}{\mathrm{e}}                               
\newcommand{\dint}[2][]{\mathop{\mathalpha{\int#1}#2}}    
\newcommand{\doint}[2][]{\mathop{\mathalpha{\oint#1}#2}}  
\newcommand{\HH}{\mathop{H\!H}\nolimits}
\newcommand{\HC}{\mathop{H\!C}\nolimits}
\newcommand{\bi}{{\bar\imath}}
\newcommand{\bj}{{\bar\jmath}}
\newcommand{\ol}[1]{\rlap{$\overline{\phantom{#1}}$}#1}
\DeclareMathSymbol{*}{\mathop}{symbols}{"03}              
\newcommand{\hoch}[3][]{\ifcase#2\relax#3\or\End(#3)\else\Hom(#3^{\otimes#2},#3)\fi^{#1}}
\newcommand{\hodge}[1]{\makebox[3.5em][c]{$h^{#1}$}}
\newcommand{\hochbox}[2]{\makebox[8em][c]{$\hoch[#2]{#1}{\CA}$}}
\newcommand{\set}[1]{\mathbb{#1}}                         
\def\C{\set{C}}                                  
\newcommand{\R}{\set{R}}
\newcommand{\Z}{\set{Z}}
\newcommand{\group}[1]{\mathop{\kern\z@\mathrm{#1}}\nolimits}     
\newcommand{\U}{\group{U}}                                
\newcommand{\opname}[1]{\mathop{\kern\z@\mathrm{#1}}\nolimits}    
\newcommand{\Tr}{\opname{Tr}}                             
\newcommand{\STr}{\opname{STr}}                           
\newcommand{\Hom}{\opname{Hom}}                           
\newcommand{\End}{\opname{End}}                           
\newcommand{\Hoch}{\opname{Hoch}}

\newcommand{\Mat}{\opname{Mat}}
\newcommand{\CA}{\mathcal{A}}

\newcommand{\CC}{\mathcal{C}}
\newcommand{\CD}{\mathcal{D}}

\newcommand{\CH}{\mathcal{H}}
\newcommand{\CL}{\mathcal{L}}
\newcommand{\CM}{\mathcal{M}}
\newcommand{\CN}{\mathcal{N}}
\newcommand{\CO}{\mathcal{O}}

\newcommand{\CS}{\mathcal{S}}
\newcommand{\CT}{\mathcal{T}}
\DeclareMathSymbol{\Bp}{\mathalpha}{boldletters}{`p}
\DeclareMathSymbol{\Bz}{\mathalpha}{boldletters}{`z}
\DeclareMathSymbol{\BG}{\mathalpha}{boldletters}{`G}
\DeclareMathSymbol{\BQ}{\mathalpha}{boldletters}{`Q}
\DeclareMathSymbol{\Bet}{\mathord}{boldletters}{"11}
\DeclareMathSymbol{\Bch}{\mathord}{boldletters}{"1F}
\ifcase \@ptsize
\or 
\or 
\fi
  \divide\@tempdima\baselineskip
  \@tempcnta=\@tempdima
\makeatother
\begin{document}
%
%

\begin{flushright}\scshape
RUNHETC-2002-04\\
hep-th/0204157\\ April 2002
\end{flushright}
\vskip15mm

\begin{center}

{\LARGE\scshape
On the Open-Closed B-Model
\par}
\vskip20mm

\textsc{Christiaan Hofman}\\[4mm]
\textit{New High Energy Theory Center, Rutgers University,\\ 
        136 Frelinghuysen Road, Piscataway, NJ 08854, USA},\\[3mm]
\texttt{hofman@physics.rutgers.edu} 

\end{center}

\vspace*{25mm}

\section*{Abstract}

We study the coupling of the closed string to the open string in the topological 
B-model. These couplings can be viewed as gauge invariant observables in the 
open string field theory, or as deformations of the differential graded algebra 
describing the OSFT. This is interpreted as an intertwining map from the closed 
string sector to the deformation (Hochschild) complex of the open string algebra.
By an explicit calculation we show that this map induces an isomorphism of 
Gerstenhaber algebras on the level of cohomology. Reversely, this can be used 
to derive the closed string from the open string. We shortly comment on 
generalizations to other models, such as the A-model. 

\newpage
\setcounter{page}{1}
%
%

\section{Introduction}
\label{sec:intro}

In this paper we will refine and work out in more detail the study 
of topological open and closed strings in \cite{homa}, focusing on 
the topological B-model. 
The main questions we will study in this paper are the coupling of 
closed string to the open string sector, and how the closed string 
is encoded in the open string. 

The question of where the closed string strings are in the open strings 
has been lingering in the context of string field theory \cite{witsft}. 
Indeed poles corresponding to the closed string operators are present 
in the open string field theory \cite{shatho1,shatho2}. 
One way is to introduce them explicitly \cite{zwie1,zwie2}, leading to 
open-closed string field theory. Another way is to find the closed string 
operators as observables in the open string \cite{grsz,hashitzh,zwie3}. 
These gauge invariant observables heuristically can be interpreted as 
integrals over a cycle of the open string field, $\Phi_C(A)=\int_CA$. 
Here $A$ is the open string field, and $C$ is induced by the closed string operator. 
In this paper we consider both approaches in a particularly simple 
topological string field theory toy model. The main idea for the latter 
approach is to view the open string field theory algebra as a noncommutative 
geometry \cite{connes}, and interpret the closed string operators 
as the cycles $C$ in this geometry. 

In \cite{homa} we discussed mixed correlators for the open/closed 
topological string; see also \cite{laza,dijk}. The open $(n+1)$-point 
functions $F_{a_0a_1\ldots a_n}$, where the $a_i$ label the open string operators, 
define structure constants for $n$-linear maps $m_n$. Together with 
the BRST operator, $Q=m_1$, they can be shown to form an $A_\infty$ algebra 
\cite{gabzwie,homa}. Especially, the 3-point functions $F_{abc}$ are the 
structure constants of a product $m=m_2$. 

The next step is to couple the boundary operators to operators in the bulk. 
This leads to mixed $(n+2)$-point functions which are defined through 
the correlators 
\begin{equation}\label{twin}
\Phi_{Ia_0a_1\ldots a_n} = 
\pm\Bigl\langle
 \hat\phi_I\hat\alpha_{a_0}\int\!\hat\alpha_{a_1}^{(1)}\cdots\int\!\hat\alpha_{a_n}^{(1)}
\Bigr\rangle,
\end{equation}
involving a bulk operator $\hat\phi_I$ and boundary operators $\hat\alpha_a$. 
Introducing deformed structure constants $F_{abc}(t)$, where $t^I$ are couplings 
for the closed string operators deforming the open string algebra, it 
follows from the Ward identities that 
\begin{equation}\label{defprod}
\del_{t^I} F_{abc} = \Phi_{Iabc}, 
\end{equation}
expressing the fact that the mixed correlators are related to the 
deformation of the algebra of boundary operators. A similar relation 
is valid for the other mixed correlators and structure constants. 
Through the mixed correlators \eqref{twin} we therefore have a natural 
map from the closed string operators to the deformations of the 
open string algebra. 

It is a well known mathematical fact that the deformation of an 
$A_\infty$ algebra is controlled by its Hochschild complex. 
This complex can be represented as the space of all multilinear 
maps on the algebra. In fact, we can interpret the mixed correlators 
\eqref{twin} as multilinear maps on the open string algebra. 
This allows us to interpret the above mentioned map as a map from 
closed string operators to the Hochschild complex. 
Both the Hochschild complex and the closed string have a structure 
of $L_\infty$ algebra. More generally, the corresponding 
cohomologies carry the structure of a Gerstenhaber algebra. 
In \cite{homa} we showed that the map from the closed string to 
the Hochschild complex intertwines these structures. Most importantly, 
the BRST operator in the closed string corresponds to the 
natural coboundary in the Hochschild complex. This implies that 
the map induces a natural map between the cohomologies. 
Moreover, this map intertwines the structure of a Gerstenhaber 
algebra. A natural question then is whether these Gerstenhaber 
algebras are actually the same. 

An example of this is given by the problem of deformation quantization 
\cite{kon1,kon2,tam,kon3,tam2}. This studies the deformation of a 
commutative product by a Poisson Lie bracket to a full 
noncommutative star-product. In \cite{cafe} it was shown that the 
solution to this problem given by Kontsevich \cite{kon1,kon2} can be 
interpreted in terms of similar correlation functions of a certain 
topological open string theory, as alluded to by Kontsevich \cite{kon1}.
The star product deformation leads 
to a noncommutative geometry \cite{connes}. In fact, this 
topological construction can be embedded in string theory, 
as was shown in \cite{scho,seiwit}, first initialized in \cite{codo}.  
A purpose of \cite{homa} was to generalize the deformations of 
\cite{cafe} to more general topological string theories. 

In this paper we want to work out in detail the relation between the 
mixed correlators and the deformations of the $A_\infty$ structure 
in more detail for the case of the B-model topological string 
\cite{witmir,kosp,bece}. 

The open string sector of the B-model is the holomorphic Chern-Simons 
theory first studied by Witten \cite{witcs}. There it was also 
shown to be a string field theory. The critical points of 
this theory are holomorphic bundles on a Calabi-Yau manifold. 
The closed string B-model is the Kodaira-Spencer theory \cite{kosp}, 
the deformation theory of complex structures. 
Its extended space of observables is given by the Dolbeault cohomology, 
or more precisely $\bigoplus_{p,q}H^{-p,q}(M)$. We find this space 
precisely as the Hochschild cohomology of the open string algebra 
$H^*_{\bdA}(M,\End(E))$. 
In other words: we find the (on-shell) closed string algebra as the 
Hochschild cohomology of the open string algebra. We want to put this 
forward as a general conjecture, which becomes:
The Hochschild cohomology of the open string field theory 
is isomorphic to the (on-shell) closed string algebra. As the 
Hochschild cohomology can be completely derived from the open string algebra, 
this implies a derivation of the closed string from the open string algebra. 
We want to see the considerations of the present paper as evidence for 
this conjecture. 

The open and closed B-model has received some attention, 
mainly in the recent year \cite{agavaf,agklva,mayr,lermay,blum}. 
This was mainly concerned with calculations of the mixed superpotential 
for non-compact Calabi-Yau manifolds. This leads in particular to 
interesting generalizations of mirror symmetry \cite{vafa,hoiqva}
to the open string sector. A different type of relation between the 
open and closed string have been studied starting from \cite{gopvaf}. 

The paper is organized as follows. 
In \secref{sec:tos} we give a short review of topological open-closed 
string theory, summarizing the results of \cite{homa}, and adding some 
general remarks on gauge invariant observables. 
In \secref{sec:bmodel} we give a brief discussion of the B-model. 

In \secref{sec:mixed-corr} we calculate the mixed correlation functions 
in the B-model with a single closed string insertion. 
In \secref{sec:def} we interpret these correlators in terms of 
deformations of the open string algebra, and present the open-closed string 
field action (the superpotential) to first order in the closed string field. 

In \secref{sec:hoch} we calculate the Hochschild cohomology of the open string 
algebra. This calculation shows that there is an isomorphism between the two. 

In \secref{sec:bvmodel} the BV structure on the Hochschild cohomology 
is used to construct a BV sigma model. We show that this model gives 
an off-shell description of the (closed) B-model. 

In \secref{sec:ncg} we interpret our results in terms of cycles in a 
noncommutative geometry. We also comment on the more precise identification 
of the closed string with the Hochschild cohomology and the relation 
to cyclic cohomology. 

In \secref{sec:other} we shortly discuss the calculations of the Hochschild 
cohomology for other models, most prominently the A-model. 

We end up with conclusion and some further discussions of our results 
in \secref{sec:concl}.

\subsection{Notation}

In this paper, the lowercase Greek indices $\mu,\nu,\ldots$ from the middle 
of the alphabet will denote holomorphic directions in the (complex) 
target space, while corresponding barred indices $\bar\mu,\bar\nu,\ldots$ denote 
the anti-holomorphic directions. In the case of open strings, these will denote 
the direction along the brane; Latin indices $i,j,\ldots$ and $\bi,\bj,\ldots$ 
will denote the (anti)holomorphic transverse directions. 
Latin indices $a,b,\ldots$ from the beginning of the alphabet will number open 
string (boundary) operators, while the uppercase Latin indices $I,J,\ldots$ from 
the middle of the alphabet number closed string operators. 
We will use hats to distinguish operators from the corresponding forms. 

The holomorphic tangent bundle of a complex manifold $M$ is denoted $\CT_M$, 
while the anti-holomorphic tangent bundle is denoted $\ol\CT_M$. 
Similarly, holomorphic and anti-holomorphic cotangent bundles are denoted 
$\CT_M^*$ and $\ol\CT_M^*$ respectively. Also $\CN_C$ will denote the 
holomorphic normal bundle of any complex cycle $C$. 
The space of sections of the 
exterior algebra of $(p,q)$-forms $\ext[p]\CT_M^*\otimes\ext[q]\ol\CT_M^*$, 
is denoted $\Omega^{p,q}(M)$. The space of $(p,q)$-forms with values 
in some (holomorphic) bundle $E$ is denoted $\Omega^{p,q}(M,E)$. 
We will call a $(0,q)$-form with values in the $p$th exterior power of the 
holomorphic tangent space $\ext[p]\CT_M$ a $(-p,q)$-form, and the space of 
such forms is denoted $\Omega^{-p,q}(M)\equiv\Omega^{0,q}(M,\ext[p]\CT_M)$. 
Analogously, we use the notation $H^{-p,q}(M)$ for its $\bar\del$-cohomology. 

The space of multilinear maps on an algebra $\CA$ (of order $n$) is 
denoted $C^n(\CA,\CA)=\Hom(\CA^{\otimes n},\CA)$. For a graded algebra 
$\CA$ we denote by $\Hom(\CA^{\otimes n},\CA)^q$ the space of $n$-linear 
maps raising the total degree by $q$. 
This is also understood as a complex with the Hochschild differential 
$\delta_m$. The cohomology of this complex is denoted $\HH^*(\CA)$.\footnote{Usually, 
the Hochschild cohomology of multilinear maps with values in a bimodule $\CM$ 
is denoted $\HH^*(\CA,\CM)$, but in this paper we only meet the case $\CM=\CA$, 
so we will not include this module in our notation.} 
The Hochschild complex of a differential graded algebra $\CA$ as a double complex 
is the same space with the two differentials $\delta_m$ and $\delta_Q$, 
and is denoted $\Hoch(\CA)$. The total cohomology of this double complex 
is denoted $H^*(\Hoch(\CA))$.

\section{Deformations of Topological Open Strings}
\label{sec:tos}

In this section we repeat the basic results of \cite{homa}, and relate 
the structure found there to gauge invariant observables from the 
point of view of string field theory.

\subsection{Correlators and Deformations}

Central in topological field theories is the existence of a BRST operator $Q$ 
such that the energy-momentum tensor $T_{\alpha\beta}$ is BRST exact. This 
implies that there must be a tensor current $b_{\alpha\beta}$ of ghost number 
$-1$ such that $\{Q,b_{\alpha\beta}\}=T_{\alpha\beta}$. The tensor $b_{\alpha\beta}$ 
is the current for a charge $G$, which is a 1-form of ghost number $-1$. 
It satisfies the anticommutation relation $\{Q,G\}=d$. The operator $G$ is 
used to define the descendants of an operator $\hat\alpha$ recursively as 
$\hat\alpha^{(p+1)}=G\hat\alpha^{(p)}$. If $Q\hat\alpha=0$, then the these are 
solutions to the descent equations $Q\hat\alpha^{(p+1)}=d\hat\alpha^{(p)}$.

Topological open strings can be characterized by the structure constants, 
which are defined by the correlation functions 
\begin{equation}\label{ainftycorr}
F_{a_0a_1\ldots a_n} = 
 (-1)^{\epsilon}\Vev{ \hat\alpha_{a_0}\hat\alpha_{a_1}\hat\alpha_{a_2}\int\!\hat\alpha_{a_3}^{(1)}\cdots\int\!\hat\alpha_{a_n}^{(1)} }. 
\end{equation}
where $\epsilon=n|\alpha_{a_1}|+\sum_{i\geq2}(n-i)|\alpha_{a_i}|$. 
When we use the open string metric $g_{ab}=\Vev{\hat\alpha_a\hat\alpha_b}$ to raise 
and lower indices, these can be interpreted as the structure constants 
of an $A_\infty$ algebra formed by multi-linear operations $m_n$, as
\begin{equation}
 m_n(\hat\alpha_{a_1},\cdots,\hat\alpha_{a_n}) = F^{a_0}_{a_1\ldots a_n}\hat\alpha_{a_0}.
\end{equation}
For example, the 3-point functions $F^a_{bc}$ are structure constants of 
the product $m=m_2$. Ward identities, similar to the WDVV equations \cite{witeqn,dvv} 
for the closed string, assure that this product is associative on-shell. 
More generally, the multi-linear operations $m_n$ satisfy higher associativity 
relations which are known as an $A_\infty$ algebra 
\cite{stas,getzjon1,getzjon2,gabzwie,zwieoc,gerst,kon4}. In this paper we will assume 
for symplicity that $m_n=0$ for $n\geq3$, so that the undeformed open string theory 
is a genuine differential graded associative algebra. This is certainly correct 
for our main focus, the B-model. 

When we include (on-shell) bulk operators $\hat\phi_I$, the mixed correlators 
can be interpreted as deformations of the $A_\infty$ algebra, by deforming 
the operations $m_n$. Introducing couplings $t^I$ for the bulk operators, 
we write the deformed structure constants as $F_{a_0a_1\ldots a_n}(t)$, 
and the deformed multilinear maps as $m_n(t)$. 
We find the following interpretation of the correlators with one bulk insertion
\begin{equation}\label{corrhoch}
\Phi_I^{(n)}(\hat\alpha_{a_1}, \ldots, \hat\alpha_{a_n}) 
 = (-1)^{\epsilon}\Vev{ \hat\phi_I\hat\alpha_b \int\!\hat\alpha_{a_1}^{(1)} \cdots \int\!\hat\alpha_{a_n}^{(1)} } g^{bc} \hat\alpha_c
 \equiv \Phi^{(n)}_{Iba_1 \ldots a_n} \hat\alpha^b 
 = \del_{t^I} F_{ba_1\ldots a_n}(0) \hat\alpha^b,
\end{equation}
where $\epsilon=|\phi_i||\alpha_{b}|+\sum_{i\geq1}(n-i)|\alpha_{a_i}|$. 
We interpret these correlators as the structure constants of a set of 
multilinear maps $\Phi_I^{(n)}$ defined by the second equality, 
which we collectively denote by $\Phi_I$. So we can write 
$\Phi_I = \sum_n\Phi_I^{(n)} = \sum_n\del_{t^I} m_n(0)$. The maps $\Phi_I$ can be 
viewed as elements of the Hochschild complex of multi-linear maps 
$C^*(\CA,\CA)=\bigoplus_n\Hom(\CA^{\otimes n},\CA)$. 

This graded space has a natural coboundary called the Hochschild differential, 
which is related to the product in $\CA$ and will be denoted $\delta_m$ 
(see e.g. \cite{kon1,kon2} and \appref{app:algebra} for a definition). 
As was shown in \cite{homa}, the BRST operator on the closed 
string operators corresponds to the Hochschild differential $\delta_m$, 
at least if we assume that the boundary operators  are taken on-shell. 
Otherwise there is a correction from the BRST operator. More precisely, the closed 
string BRST operator acting on $\hat\phi_I$ corresponds to $\delta_m+\delta_Q$ 
acting on $\Phi_I$, where $\delta_Q$ is the supercommutator with the open string 
BRST operator $Q$, where the action on several boundary operators is interpreted 
appropriately. More generally, when other structure constants of the $A_\infty$ 
algebra are nonzero, we find that the closed string BRST operator corresponds to 
$\sum_n \delta_{m_n}$, where $\delta_{m_n}$ is defined analogously 
to the Hochschild differential, using the canonical Gerstenhaber bracket 
$[\cdot,\cdot]$ on the Hochschild complex. More precise definitions of the 
Gerstenhaber bracket and the Hochschild differential can be found in \appref{app:algebra}.

\subsection{Deformations of dg-Algebras and Gerstenhaber Algebras}

In the case of a dg-algebra, that is we only have $Q=m_1$ and an associative 
product $m=m_2$, we find that the Hochschild complex is a double complex, 
with differentials $\delta_Q$ and $\delta_m$. In the rest of this paper 
we will assume that the undeformed open string theory has this structure 
of a dg-algebra. The map from closed string operators $\phi_I$ to the 
multilinear operations $\Phi_I$ through the correlation functions naturally 
relates the closed string operators to the Hochschild complex. 
The closed string BRST operator corresponds to the total Hochschild coboundary 
$\delta_m+\delta_Q$ on this double complex under this map. 
It follows that the on-shell closed string algebra, that is the BRST cohomology, 
is naturally related to the Hochschild cohomology $H^*(\Hoch(\CA))$, 
which is the total cohomology of this double complex. 
To calculate this total cohomology, we can use a spectral sequence calculation. 
To calculate the first term in the spectral sequence, one can make 
use of the Hochschild-Kostant-Rosenberg theorem, or the analytic generalization 
due to Connes. This expresses the cohomology with respect to $\delta_m$ of 
the Hochschild complex of a polynomial algebra as a polynomial algebra. 
On the first term of the spectral sequence we now have the coboundary $\delta_Q$ 
(or more precisely the induced coboundary in the Hochschild cohomology). 
The corresponding cohomology in turn constitutes the second term in the 
spectral sequence. In general, the spectral sequence does not have to 
terminate here. Generally, we still have a remnant of $\delta_m$, if it is 
not completely compatible with the cohomology of $\delta_Q$, in the sense 
that we can not take homogeneous representatives which are simultaneously 
closed under both coboundaries. We will see that this indeed happens for HCS.
This gives rise to descent equations of the form 
$\delta_m\Phi^{(n)} = -\delta_Q\Phi^{(n+1)}$. Solving these descent equations, 
we see that the total sum $\Phi=\sum_n\Phi^{(n)}$ is indeed closed with respect 
to the total coboundary, $(\delta_m+\delta_Q)\Phi=0$. 

Closed string operators form naturally a very interesting algebraic structure 
see e.g. \cite{wizwi,zwie,stasheff,kvz,ksv,liz}. 
Well known is the (OPE) product on the closed string. It is given by the constant 
term in the OPE. On shell, this product is associative and symmetric. 
Another structure is the bracket, which is found in terms of the contour integral 
of one operator around each other. In conformal gauge this is the residue, 
the coefficient of the $1/z$ pole, of the OPE. 
It is related to the current algebra of the closed string. On-shell, these 
two operations satisfy the relations of what is called a Gerstenhaber algebra. This 
is a graded version of a Poisson algebra, for which the bracket has degree $-1$. 

The Hochschild cohomology of an associative algebra is also known to have the 
structure of a Gerstenhaber algebra. The symmetric product is defined by the 
so-called cup product, denoted $\cup$ and defined in \appref{app:algebra}, 
and the bracket is defined by a graded version of composition of multilinear maps. 
It turns out that for topological strings the map defined by the mixed correlators 
intertwines the structure of Gerstenhaber algebra of the closed string and the 
Hochschild cohomology of the open string algebra. In more concrete terms the 
product $\hat\phi_I\cdot\hat\phi_J$ of two closed string operators corresponds 
to the map $\Phi_I\cup\Phi_J$, while the bracket of a pair of operators 
$\{\hat\phi_I,\hat\phi_J\}$ corresponds to the Gerstenhaber bracket 
of the corresponding maps $[\Phi_I,\Phi_J]$ in the Hochschild complex.

\subsection{Open String Field Theory Action and Gauge Invariant Observables}

We can make contact between the above construction and 
open string field theory. We saw that in general the topological open 
string algebra $\CA$ has the structure of an $A_\infty$ algebra. 
We assumed the undeformed case to be a graded differential associative 
algebra, given by the BRST operator $Q$ and the product $m$. 
This can be understood in terms of an open string field theory. 
The open string field is expanded in the boundary operators as  
$A=A^a\hat\alpha_a$. We will here reduce only to the degree 
1 part of the string field. In the explicit case of the D-brane this 
corresponds to the 1-form gauge field, and gives the physical 
part of the field. In principle also other degrees could be included, 
which are interpreted as ghosts and anti-ghosts. 
In order to write down the action we need an inner product 
$\vev{\cdot,\cdot}$ on the algebra $\CA$, which is provided by the 
2-point functions. The algebraic structures $Q$ and $m$ will be 
cyclic using this inner product. 
The action can then be written as \cite{witsft,witcs}
\begin{equation}
  S_0 = \frac{1}{2}\Vev{A,QA} + \frac{1}{3}\Vev{A,m(A,A)}.
\end{equation}
Often the inner product is written as a (formal) integral. 
This action has a gauge invariance by 
\begin{equation}
  \delta_\Lambda^0 A = Q\Lambda+m(A,\Lambda)-m(\Lambda,A), 
\end{equation}
where $\Lambda$ is a degree zero field. 

We saw above that we can understand the multilinear operations 
$\Phi$ as deformations of the $A_\infty$ structure constants. 
In this they will also deform the string field theory action 
above. We can alternatively understand them as observables in 
the open string field theory. Indeed observables can be used 
to deform the action by exponentiating them. An important criterion 
for these observables is that they respect the gauge invariance. 
We will now see that indeed, in an appropriate sense, they do. 

Let us first consider the case of a closed $\Phi$ having only a 
component $\Phi^{(0)}$ of order 0. This implies that it is closed 
with respect to both coboundaries, $\delta_Q\Phi^{(0)}=0=\delta_m\Phi^{(0)}$. 
As $\Phi^{(0)}\in C^0(\CA,\CA)=\CA$, we can define an observable 
\begin{equation}
  \CO_\Phi = \Vev{A,\Phi^{(0)}}. 
\end{equation}
This can be heuristically considered as an ``integral'' of the ``gauge field'' 
$A$ over a ``cycle'' given by $\Phi^{(0)}$. In the analogue of string field theory 
with noncommutative geometry \cite{witsft}, it is natural to relate the 
multilinear maps $\Phi$ with cycles. We saw they are elements of the 
Hochschild cohomology, while in noncommutative geometry the latter 
is related to the cycles in the noncommutative space \cite{connes}. 
To see the gauge invariance of this observable we note 
the following two consequences of the above restriction on $\Phi^{(0)}$. 
From $\delta_Q\Phi^{(0)}=0$ we have 
\begin{equation}
  \Vev{Q\Lambda,\Phi^{(0)}} = -\Vev{\Lambda,\delta_Q\Phi^{(0)}} = 0. 
\end{equation}
The constraint $\delta_m\Phi^{(0)}=0$ implies 
\begin{equation}
  \Vev{m(A,\Lambda)-m(\Lambda,A),\Phi^{(0)}} = \Vev{\Lambda,\delta_m\Phi^{(0)}(A)} = 0.
\end{equation}
Combining these, we find that $\CO_\Phi$ is indeed gauge invariant. 

Let us now look at a more general case, where $\Phi$ has components of 
any order. We still take $\Phi$ to be closed; so the components satisfy 
$\delta_m\Phi^{(n)}+\delta_Q\Phi^{(n+1)}=0$. Also the 
maps $\Phi^{(n)}$ can be shown to be cyclic. 
We take the following ansatz for the corresponding observable 
\begin{equation}\label{observ}
  \CO_\Phi = \sum_{n\geq 0}\frac{(-1)^{\frac12(n-1)(n-2)}}{n+1}\Vev{A,\Phi^{(n)}(A,\cdots,A)}. 
\end{equation}
We can then derive the following expression for the variation of the 
deformed action under the gauge transformations 
\begin{equation}
  \delta_\Lambda^0 \CO_\Phi = \sum_{n\geq 1}\sum _{i=1}^{n}(-1)^{\frac12(n-1)(n-2)+i}
   \Vev{F,\Phi^{(n)}(A,\cdots,A,\mathop{\Lambda\kern0pt}_i,A,\cdots,A)}.
\end{equation}
where the $i$ indicates that $\Lambda$ is inserted at the $i$th place, 
and $F=QA+m(A,A)$ is the ``field strength'' of $A$. 
As $F=\frac{\delta S_0}{\delta A}\approx 0$ is the equation of motion 
of the undeformed theory, we find gauge invariance on shell, at least to 
first order in the deformation. This observation can be used to find full 
gauge invariance to first order. We modify the gauge transformation law 
by terms involving $\Phi$, 
\begin{equation}\label{defgauge}
  \delta_\Lambda' A = \sum_{n\geq 1} \sum_{i=1}^n(-1)^{\frac12n(n+1)+i}
  \Phi^{(n)}(A,\cdots,A,\mathop{\Lambda\kern0pt}_i,A,\cdots,A).
\end{equation}
Then the variation of the undeformed action $S_0$ will exactly cancel the 
deviation form gauge invariance of $\CO_\Phi$ above. Notice that this is quite 
natural, as $\Phi^{(n)}$ are the deformation of the $A_{\infty}$ structure. 
So although the $\CO_\Phi$ are now not genuinely gauge invariant expressions, 
the total expression $S_0+\CO_\Phi$ is gauge invariant under the modified 
gauge transformations $\delta_\Lambda^0+\delta_\lambda'$, to first order in $\phi$. 

As $\CO_\Phi$ transforms nontrivially under the modification $\delta_\Lambda'$ 
the combination $S_0+\CO_\Phi$ is not exactly gauge invariant with respect 
to the modified gauge transformations. The corrections are however 
higher order in the deformation $\phi$. We can generalize the gauge 
invariance to higher orders; we should then however also take into account 
the higher order corrections to the maps $\Phi$. We therefore replace them 
by the completely deformed multilinear operations $\bar\Phi$. They can be 
derived from the fully deformed correlation functions, and should satisfy 
the master equation 
\begin{equation}\label{master}
  \delta_Q\bar\Phi^{(n)} + \delta_m\bar\Phi^{(n-1)} 
  + \sum_{k=0}^{n}(-1)^{(n-1)k}\bar\Phi^{(n+1-k)}\circ\bar\Phi^{(k)} = 0. 
\end{equation}
In \cite{homa} it was shown that this is satisfied by the 
mixed correlators completely deformed by a closed string operator 
$\phi$ provided it is BRST closed and it satisfies the additional 
identity $\{\phi,\phi\}=0$. 

The gauge deformation is deformed by $\delta_\Lambda'$ given by the same 
formula \eqref{defgauge} replacing $\Phi$ by $\bar\Phi$, and 
$\delta_\Lambda' S_0=\Vev{F,\delta_\Lambda'A}$. 
The variation of the observable under a gauge transformation can be written as 
\begin{equation}
  \delta_\Lambda\CO_{\bar\Phi} = \sum_n (-1)^{\frac12(n-1)(n-2)}
  \Vev{\delta_\Lambda A,\bar\Phi^{(n)}(A,\cdots,A)}.
\end{equation}
Using the cyclicity of the correlation functions $\bar\Phi$, we can derive 
the following identities 
\begin{eqnarray}
  \Vev{\Lambda,\delta_Q\bar\Phi^{(n)}(A^n)} &=& -\Vev{Q\Lambda,\bar\Phi^{(n)}(A^n)} 
  + \sum_{k}(-1)^{n+k}\Vev{QA,\bar\Phi^{(n)}(A^{k},\Lambda,A^{n-k-1})},\\
  \Vev{\Lambda,\delta_m\bar\Phi^{(n-1)}(A^{n})} &=& (-1)^{n-1}\Vev{m(A,\Lambda)-m(\Lambda,A),\bar\Phi^{(n-1)}(A^{n-1})} \nonumber\\
  &&+ \sum_{k}(-1)^{k+1}\Vev{m(A^2),\bar\Phi^{(n-1)}(A^{k},\Lambda,A^{n-k-2})},\\
  \Vev{\Lambda,\bar\Phi^{(n+1-l)}\circ\bar\Phi^{(l)}(A^{n})} &=& 
  \sum_{k}(-1)^{n-k-l}\Vev{\bar\Phi^{(l)}(A^{l}),\bar\Phi^{(n+1-l)}(A^{k},\Lambda,A^{n-k-l})}.
\end{eqnarray}
Here $A^n$ stands for the $n$ times repeated arguments $A$. 
The sum of the left hand sides is the master equation, and therefore 
vanishes by our assumption. Summing over the right hand sides including 
a proper sign, one finds they sum up to 
$\delta_\Lambda^0\CO_{\bar\Phi}+\delta_\Lambda' S_0+\delta_\Lambda'\CO_{\bar\Phi}$. 
This shows that the deformed action $S_0+\CO_{\bar\Phi}$ 
is gauge invariant for the full gauge transformation $\delta_\Lambda^0+\delta_\Lambda'$. 

Having identified the closed string operators with the elements of the 
Hochschild cohomology, we can now invert the construction of gauge 
invariant operators. That is, for any element $\Phi$ of the Hochschild 
cohomology of the open string algebra $\CA$ we construct the above 
operator $S$. The descent equation satisfied by the cohomology classes 
guarantee that this is gauge invariant at least to first order in the 
deformed gauge invariance.

\section{The B-Model}
\label{sec:bmodel}

We now discuss in some detail the topological open string related to Kodaira-Spencer theory, 
which is described by holomorphic Chern-Simons \cite{witcs}. This topological 
string is also known as the B-model. 
A more detailed discussion of the B-model can be found in \cite{witmir}.

\subsection{Action and BRST Symmetry}

We consider a Calabi-Yau space $M$. In general, we could also consider
a general complex manifold, but for non-Calabi-Yau spaces there is an
anomaly which makes it hard to define a physical theory (the theory is
not unitary anymore). 

The topological B-model is a twisted version of the $(2,2)$ supersymmetric 
CFT with target space $M$. The fields of this topological sigma-model are 
the bosonic coordinate fields $z^\mu,\bar z^{\bar\mu}$, two sets of 
twisted fermions $\bar\eta^{\bar\mu},\chi_\mu$ transforming as worldsheet scalars, 
and a set of twisted fermions $\rho^\mu$ transforming as worldsheet 1-forms. 
The action is given by 
\begin{equation}\label{baction}
  S = t\int_\Sigma \Bigl( g_{\mu\bar\mu}dz^\mu*d\bar z^{\bar\mu} 
   - g_{\mu\bar\mu}\rho^\mu *D\bar\eta^{\bar\mu}\Bigr)
   + \frac{1}{\kappa}\int_\Sigma \Bigl(
      \rho^\mu D\chi_\mu  
   - \frac{1}{2} R^\lambda_{\mu\bar\mu\nu}\rho^\mu\rho^\nu\bar\eta^{\bar\mu}\chi_\lambda\Bigr),
\end{equation}
where $R^\lambda_{\mu\bar\mu\nu}=g^{\lambda\bar\lambda}R_{\bar\lambda\mu\bar\mu\nu}$ 
is the curvature of the K\"ahler manifold $M$ and $D$ is a covariantized derivative. 
In the action we have introduced an extra coupling parameter. 
Usually the two coupling parameters are identified as $t=\frac{1}{\kappa}$. 
For our purpose it will be more useful to leave them as independent parameters. 

This action has a BRST symmetry $Q$ given by 
\begin{eqnarray}
  Q\bar z^{\bar\mu} = \bar\eta^{\bar\mu},\qquad Q\rho^\mu = dz^\mu,\qquad
  Qz^\mu = Q\bar\eta^{\bar\mu} = Q\chi_\mu = 0.
\end{eqnarray}
Identifying the $\bar\eta^{\bar\mu}$ with the 1-forms $d\bar z^{\bar\mu}$ on $M$, 
we can identify $Q$ with the Dolbeault differential $\bar\del$ on $M$. 
The action \eqref{baction} can be written as the sum of a topological term 
and a BRST exact term, as 
\begin{equation}
  S = \frac{1}{\kappa}\int_\Sigma\rho^\mu d\chi_\mu 
   + Q\int_\Sigma\Bigl( tg_{\mu\bar\mu}\rho^\mu *d\bar z^{\bar\mu} 
   - \frac{1}{2\kappa}\Gamma^\lambda_{\mu\nu}\rho^\mu\rho^\nu\chi_\lambda\Bigr). 
\end{equation}
As all the $t$-dependence is in the exact terms, this shows that the 
B-model is independent of this parameter. Similarly, we see from this that 
the B-model does not depend on the K\"ahler class of the metric $g_{\mu\bar\mu}$ 
and on the worldsheet metric. It depends only on the 
complex structure of $M$. As $t$ can be identified with an inverse coupling 
constant, this also means that the lowest order weak coupling expansion is exact. 
This will turn out very useful in calculating correlation functions, 
as we can safely take the limit $t\to\infty$. 
Because the last term in the action is only BRST closed, the B-model potentially 
does depend on the coupling $\kappa$. Although this dependence was somehow suppressed 
in the pure open string case \cite{witcs}, we will see that this is not 
necessarily true for the coupling between the open and the closed sector. 
The dependence will however be rather mild and can mostly be undone 
by a rescaling. There will be a crucial difference between vanishing 
and nonvanishing values of $\kappa$.
 
On closed worldsheets there exists a second BRST operator $Q'$ given by 
\begin{equation}
  Q'\bar z^{\bar\mu} = \frac{g^{\bar\mu\mu}}{\kappa t}\chi_\mu,\qquad
  Q'\rho^\mu = *dz^\mu,\qquad
  Q'z^\mu = Q'\chi_\mu = 0,\qquad
  Q'\bar\eta^{\bar\mu} = -\del_{\bar\nu}g^{\mu\bar\mu}\bar\eta^{\bar\nu}\chi_\mu,
\end{equation}
but it will not play a very important role in the present paper. 
It will be explicitly broken by the boundary conditions, as $\bar z^{\bar\mu}$ 
and $\chi_\mu$ will never have the same boundary condition. 
It corresponds to the Dolbeault operator $\del^\dagger$. 

It can be shown that indeed the energy momentum tensor is BRST exact, 
$T_{\alpha\beta}=\{Q,b_{\alpha\beta}\}$, showing that the theory is indeed 
topological \cite{witcs}. The 1-form charge $G$ for this current 
has the following action on the fields 
\begin{equation}
\begin{array}{r@{\;}c@{\;}l@{\qquad}r@{\;}c@{\;}l}
  Gz^\mu &=& \rho^\mu,\qquad\hfil
  G\bar z^{\bar\mu} = 0,\hfil&
  G\bar\eta^{\bar\mu} &=& d\bar z^{\bar\mu}, \\
  G\chi_\mu &=& \kappa tg_{\mu\bar\mu}*d\bar z^{\bar\mu}
   -\Gamma^\lambda_{\mu\nu}\rho^\nu\chi_\lambda,&
  G\rho^\mu &=& -\displaystyle \frac{1}{2}\Gamma^\mu_{\nu\lambda}\rho^\nu\rho^\lambda. 
\end{array}
\end{equation}
It can be straightforwardly checked that there is an on-shell relation 
$\{Q,G\}\approx d$, reflecting the above relation 
between the currents $b$ and $T$. This operator can be considered as the target 
space operators $\bar\del^\dagger$ and $\del$. The two BRST operators $Q,Q'$ 
and the two components $G$ are related by a twist to the the four supersymmetry 
operators of the $(2,2)$ theory. The operator $Q'$ can be shown to satisfies 
the analogous relation $\{Q',G\}\approx *d$. Therefore we have also 
$T_{\alpha\beta}=\{Q',b'_{\alpha\beta}\}$, where $b'_{\alpha\beta}$ is related 
to $b_{\alpha\beta}$ by a Hodge duality on one of the indices --- it is the 
current for $G'=-*G$.

\subsection{Kodaira-Spencer Theory}

The closed string B-model \cite{witmir} is governed by the Kodaira-Spencer 
theory \cite{kosp}, which studies the moduli space of complex structures. 
KS theory is known quite well; many details from the point of view of the 
B-model can be found for example in \cite{bece}. 

The operators of the closed B-model are build using the fermionic scalars 
$\chi_\mu$ and $\bar\eta^{\bar\mu}$. The operator corresponding to a form 
have the form \cite{witmir} 
\begin{equation}\label{clop}
\hat\phi = \phi^{\mu_1\ldots \mu_p}_{\bar\mu_1\ldots\bar\mu_q}(z,\bar z)
 \bar\eta^{\bar\mu_1}\cdots\bar\eta^{\bar\mu_q}\chi_{\mu_1}\cdots\chi_{\mu_p},
\end{equation}
where $\phi\in \Omega^{0,q}(M,\ext[p]\CT_M)$ is a $(0,p)$-form with values 
in the exterior powers of the holomorphic tangent space. 
We see that the action of the BRST operator $Q$ indeed correspond to 
the operator $\bar\del$ on these forms. Therefore the BRST closed operators 
correspond to the closed forms $\phi$, and the BRST cohomology of the 
closed string can be identified with the space of vector valued forms on the 
complex space $M$, 
\begin{equation}\label{cplxmod}
\bigoplus_{p,q}H^{-p,q}(M) \equiv \bigoplus_{p,q}H^q_{\bar\del}(M,\ext[p]\CT_M) 
\cong \bigoplus_{p,q}H^{3-p,q}(M),
\end{equation}
where $\CT_M$ denotes the holomorphic tangent space to $M$, 
and we take the cohomology with respect to $\bar\del$. 
The last equivalence is given by contraction with the holomorphic 3-form $\Omega$. 

The physical operators are the closed string operators of total ghost number 2, 
that is operators for which $p+q=2$. Usually one considers Calabi-Yau's
having $h^{1,0}=0$. This implies that the only physical operators 
are in $H^{-1,1}(M)$. They correspond to the deformation of the complex structure. 
Here we will also consider more general ``CY's'' for which the above is not 
necessarily true, so may have even more reduced holonomy. 
The 3-point function on the sphere of three physical operators 
$\hat\phi_I$, $\hat\phi_J$ and $\hat\phi_K$ is given by \cite{witmir}
\begin{equation}
 \Vev{\hat\phi_I\hat\phi_J\hat\phi_K}_{S^2} 
  = \int\Omega\wedge\phi_I\cdot\phi_J\cdot\phi_K\cdot\Omega. 
\end{equation}
Here the holomorphic vector indices are first fully contracted with $\Omega$. 
Note that there is a selection rule: this is only nonzero when 
$\sum_I p_I=\sum_I q_I=3$. This is of course nothing but the conservation 
of left and right ghost number, which are known to have an anomaly of 3. 
Also this is satisfied when all 3 are in $H^{-1,1}(M)$. Similarly, 
when one considers allows for non-physical operators, having different 
charges, one has obvious $n$-point functions when the above selection 
rule is satisfied.

\subsection{Holomorphic Chern-Simons Theory}

The open string sector of the B-model is given by 
holomorphic Chern-Simons theory (HCS) \cite{witcs}. The KS theory 
couples naturally to holomorphic bundles, or more generally, 
even dimensional D-branes wrapped around holomorphic cycles in 
the Calabi-Yau with a holomorphic bundle on it. These are the natural 
boundary conditions which in fact are invariant under the BRST operator 
of KS. For simplicity, we will mainly talk about
bound states containing a nonzero number of D-branes fully wrapped around
the Calabi-Yau. The D-brane system is then described by a holomorphic 
bundle $E$ over the Calabi-Yau.
The tangent space of the model, spanned by the boundary operators 
$\alpha_a$, are the deformations of the holomorphic connection 
$A$ of the holomorphic bundle $E$ on the Calabi-Yau. The BRST operator 
of the open string becomes the covariantized form of the closed string 
(total) BRST operator, $Q_B = \bdA$. The derivation condition $Q^2=0$ 
translates to $F^{0,2}=\bdA^2=0$, expressing the fact that the 
bundle has to be holomorphic. The tangent space is at any point isomorphic 
to $\Omega^{0,1}(M,\End(E))$, the space of $(0,1)$-forms with values 
in the adjoint bundle. We will simply denote the variations by $\alpha$. 
The corresponding zero-form vertex operator and its descendant are given 
by 
\begin{equation}
  \hat\alpha = \alpha_{\bar\mu}(z,\bar z)\bar\eta^{\bar\mu},\qquad
  \hat\alpha^{(1)} = \alpha_{\bar\mu}(z,\bar z)d\bar z^{\bar\mu}
   + \del_\mu\alpha_{\bar\mu}(z,\bar z)\rho^\mu\bar\eta^{\bar\mu},
\end{equation}
where the $\sigma$ is a coordinate along the boundary. The fermions 
$\chi_\mu$ do not appear in the operators as they satisfy Dirichlet boundary 
conditions. The BRST cohomology of the open string B-model can therefore be 
identified with 
\begin{equation}
H^*_{\bdA}(M,\End(E)).
\end{equation}

Using this correspondence of operators and (variation of the) gauge field, 
we can  express all  correlators of HCS in terms of an effective field theory, 
living on the worldvolume of the D-brane. 
The effective action is given by \cite{witcs}
\begin{equation}
S_0 = \int_M \Omega \wedge\Tr
\biggl(\frac{1}{2}A\wedge\bar\del A + \frac{1}{3}A\wedge A\wedge A\biggr).
\end{equation}
For example, the structure constants of the open string algebra are 
given in terms of this effective field theory by 
\begin{equation}
F_{abc} = \int_M \Omega \wedge\Tr
\Bigl(\alpha_a\wedge\alpha_b\wedge\alpha_c \Bigr).
\end{equation}

In other words, the theory is given by the cubic OSFT of Witten 
\cite{witsft}, with BRST operator $\bdA$, product $\wedge$ and integral/trace 
$\int\Omega\wedge\Tr$.

This discussion can be generalized for B-branes wrapping a holomorphic 
cycle $C\subset M$. When $C$ is a genuine submanifold of $M$, there are 
extra scalars $X$ in the B-model. They can be considered as describing the 
transverse motion of the brane. They are therefore sections of the 
holomorphic normal bundle to $C$, which we will denote $\CN_C$. These scalar fields 
transform in the adjoint of the gauge group. Hence, they will actually give rise 
to a field $X\in \Omega^{0}(C,\CN_C\otimes\End(E))$. In the following, 
indices $\mu,\nu,\ldots$ and the corresponding antiholomorphic indices 
will be used for directions along $C$, and indices $i,j,\ldots$ for 
direction in the holomorphic normal bundle $\CN_C$. Therefore, the scalar fields 
will have coordinates $X^i$. Perturbations of the scalar fields will give rise to 
operators in the B-model. To distinguish them from the 1-form operators, we will 
denote them $\gamma$. 
The space of on-shell operators (including the ghost and anti-ghost sectors) 
corresponds to the cohomology $H^*_{\bdA}(C,\ext[*]\CN_C\otimes\End(E))$, 
with the physical operators in degree 0. The boundary operator corresponding to 
an element $\gamma\in H^0(C,\CN_C\otimes\End(E))$ is given by 
\begin{equation}
  \hat\gamma = \gamma^i\chi_i,\qquad
  \hat\gamma^{(1)} = \kappa t\gamma^ig_{i\bi}*d\bar z^{\bi}
   - \gamma^i\Gamma^j_{ki}\rho^k\chi_j + \del_\mu\gamma^i\rho^\mu\chi_i,
\end{equation}
where the subscript $n$ denotes the normal direction to the boundary. 
In order for this to be nontrivial on the boundary, the fields $\chi_i$ should 
satisfy Neumann boundary conditions. Furthermore, the fields $z^i$, $\bar z^\bi$, 
and $\bar\eta^\bi$ will satisfy Dirichlet boundary conditions. 

Let us first assume that $C$ is a point. In this case there are 
three zero-modes for the fermions $\chi_i$ which have to be provided by as 
many scalar fields. More scalar fields would lead to contractions which vanish 
when we take $t\to\infty$. Therefore, the complete contribution comes 
from the zero modes, giving rise to a cubic interaction. The action is 
\begin{equation}
  S_0 = \frac{1}{3}\Omega\cdot\Tr\Bigl(XXX\Bigr), 
\end{equation}
where the holomorphic 3-form $\Omega$ is evaluated at the point $C$ and 
fully contracted with the vector indices of the field $X$. 
Next consider the case where $C$ is a complex curve in $M$. Then there are 
two zero-modes for the normal fermions $\chi_i$ and a single one for 
$\bar\eta^{\bar\mu}$. Reducing to zero-mode integrals, again noting 
that contractions are subleading, we find the action 
\begin{equation}
  S_0 = \int_C\Omega\cdot\Tr\biggl(\frac{1}{2}X\bar\del X+XAX\biggr).  
\end{equation}
The first term is induced by the BRST operator. 
When $C$ a complex surface, there is one zero-mode 
for $\chi_i$ and two for $\bar\eta^{\bar\mu}$. The action is then 
\begin{equation}
  S_0 = \int_C\Omega\cdot\Tr\Bigl(X\bar\del A+XA\wedge A\Bigr). 
\end{equation}
Note that in all three cases the action can be found from dimensional 
reduction.

\section{Mixed Correlators in the Open-Closed B-Model}
\label{sec:mixed-corr}

We now turn to our main objective, which is the calculation of 
correlators in the open-closed B-model.

\subsection{Mixed Correlators}

We will calculate the 
mixed correlators, involving a single closed string operator, for a 
brane wrapped completely around $M$. Therefore, there are only open string 
operators $\hat\alpha$, corresponding to the gauge field. 

To calculate the mixed correlators we need the propagators between the various 
$\beta_I\in H^{0,2}(M)$. There are no possible contractions in the $t\to\infty$ 
limit, and the correlators is completely determined by the zero-mode integral. 
As there are three zero-modes for $\bar\eta^{\bar\mu}$ we need a single 
boundary insertion, giving a mixed 2-point function 
\begin{equation}\label{def02}
  \Bigl\langle \hat\beta_I\hat\alpha_a\Bigr\rangle 
 = \int \beta_I\wedge\Omega\wedge \Tr(\alpha_a).
\end{equation}

Next we consider a closed string operator corresponding to $\varphi\in H^{-1,1}(M)$. 
As this contains a $\chi_\mu$, which has no zero-modes, the 2-point function vanishes. 
The relevant correlator is a mixed 3-point function 
$\Vev{\hat\varphi_I\hat\alpha_a\dint{\hat\alpha_b^{(1)}}}$. 
Writing out the operators we have 
\begin{equation}\label{def20}
  \dint{d\sigma}\Vev{({\varphi_I}_{\bar\mu}^\mu\bar\eta^{\bar\mu}\chi_\mu)(u)(\alpha_{a\bar\lambda}\bar\eta^{\bar\lambda})(0)(\alpha_{b\bar\nu}\del_\sigma{\bar z}^{\bar\nu}+\del_\nu\alpha_{b\bar\nu}\rho_\sigma^\nu\bar\eta^{\bar\nu})(\sigma)}. 
\end{equation}
Because $\chi_\mu$ has no zero-modes, the only potential contributions come 
from contractions of this field. Therefore we need a fermion $\rho^\nu$ somewhere 
in the expression. This implies that we need a first descendant on the boundary, 
for which we need at least two open string operators. The contraction of 
$\chi_\mu$ in $\hat\varphi_I$ with $\rho^\mu$ in $\hat\alpha_b$ will give 
a propagator $\sim\frac{\kappa}{\sigma-u}$. When we introduce more boundary operators in the 
correlator (as first descendants), they will not contribute as $\rho$ 
and $\del_\sigma{\bar z}$ have no zero-mode and further contractions 
always give $1/t$ contributions, which vanish for $t\to\infty$. 
We conclude that the correlator is given by 
\begin{equation}\label{def11}
\Bigl\langle \hat\varphi_I\hat\alpha_a\dint{\hat\alpha_b^{(1)}}\Bigr\rangle 
 = \kappa\int \varphi_I\cdot\Omega\wedge \Tr(\alpha_a\wedge\del\alpha_b)
\end{equation}

We should note at this point that there are potential contributions 
from the boundary of the integration, when two open string operators collide. 
The contribution will however always involve a $\bar\eta\rho$ or $z\bar z$ 
contraction, and therefore give contributions of order $1/t$, which would 
vanish in our limit. Though the integrals are potentially divergent, 
we will regularize them using a point-splitting procedure, thereby 
avoiding any of these contributions. We will come back to this point in more 
detail later. 

For $\theta\in H^{-2,0}(M)$ the result is 
\begin{equation}
  \Bigl\langle \hat\theta_I\hat\alpha_a\dint{\hat\alpha_b^{(1)}}\dint{\hat\alpha_c^{(1)}}\Bigr\rangle 
 = \frac{\kappa^2}{2}\int \theta_I\cdot\Omega\wedge \Tr(\alpha_a\wedge\del\alpha_b\wedge\del\alpha_c). 
\end{equation}
The calculation is similar to the one above. Only now $\hat\theta_I$ contains two 
fermions $\chi_\mu$ which we need to contract. Therefore, we need two 
descendants, and therefore at least three boundary operators. The 
contractions of the two $\chi_\mu$ factors in $\hat\theta_I$ with the 
$\rho^\mu$ in the two descendant operators give a nonzero integral 
proportional to the residue 
of the OPE. The integral over the insertions of $\hat\alpha_b$ and $\hat\alpha_c$ 
can be written as angular integrals, where we have take into account of 
the order. This will give an integral over a 2-simplex, which 
is responsible for the factor of $\frac{1}{2}$ in front of the result. 
We note that in general when there are $n$ first descendants integrated in 
fixed order along the boundary, this will be an integral over an 
$n$-simplex, giving a factor of $\frac{1}{n!}$. 
More insertions of boundary operators will again give zero, 
as more contractions give subleading corrections in $1/t$, and therefore 
these are the only nonzero correlators involving a single $\hat\theta_I$.

\subsection{Including Scalars}

Let us now generalize to the case where the B-brane wraps a holomorphic 
cycle $C\subset M$. We will not discuss this in all detail, 
but rather look a few examples. With what we have learned so far, this 
will allow us to find the general rule to construct the interactions. 
We take for $C$ a complex curve in $M$, and denote the single complex 
coordinate $z^\mu$ by $z$. First, we take a closed string operator 
corresponding to a complex structure deformation of the form
$\hat\phi_I=\varphi_I{}_{\bar z}^i\bar\eta^{\bar z}\chi_i$, and a single open 
string operator. We can reduce completely to zero-mode integrals, giving 
\begin{equation}
  \Vev{\hat\phi_I\hat\gamma_a} = \Vev{(\varphi_I{}_{\bar z}^i\bar\eta^{\bar z}\chi_i)(\gamma_a^j\chi_j)} 
  = \dint[_C]{dz\wedge d\bar z}\varphi_I{}_{\bar z}^i\Omega_{zij}\Tr(\gamma_a^j). 
\end{equation}
For more open string insertions, we insert descendants of the boundary operators. 
The descendants contain two terms. Let us consider the mixed 3-point function 
$\Vev{\hat\phi_I\hat\gamma_a\dint{\hat\gamma_b^{(1)}}}$. The main term contributing will be 
\begin{equation}
 \kappa t\Vev{(\varphi_I{}_{\bar z}^i\bar\eta^{\bar z}\chi_z)(\gamma_a^j\chi_j)(\gamma_b^kg_{k\bi}\del_n\bar z^\bi)} 
  = \kappa \dint[_C]{dz\wedge d\bar z}\del_k\varphi_I{}_{\bar z}^i\Omega_{zij}\Tr(\gamma_a^j\gamma_b^k).
\end{equation}
Here we used a contraction between $z^k$ and $\del_n\bar z^\bi$. Notice that 
this would naively vanish in the $t\to\infty$ limit as the propagator is of 
order $1/t$. However this is precisely canceled by the explicit factor of 
$t$ in front of the correlator (coming from the descent procedure). 
Furthermore the pure zero-mode contribution vanishes, which makes the 
$t\to\infty$ limit defined.
The other term of the descendant will not contribute as there are no zero-modes 
or contractions possible. There are higher point functions involving extra 
insertion of scalar operators, which give similar contributions. I.e, any further 
operator $\hat\gamma_c$ in the correlator gives an insertion of $\gamma_c^l\del_l$, 
where the derivative acts on $\varphi$. 

Next, we take a complex structure deformation of the form 
$\hat\phi_I=\varphi_I{}_{\bar z}^z\bar\eta^{\bar z}\chi_z$. 
As this does not contain any $\chi_i$, we need at least two 
open string fields to soak up the corresponding zero-modes. 
One of these is a descendant. The $\chi_z$ has to contract 
with a $\rho^z$ in the descendant operator. This gives 
\begin{equation}
  \Vev{\hat\phi_I\hat\gamma_a\dint{\hat\gamma_b^{(1)}}} 
  = \kappa \Vev{(\varphi_I{}_{\bar z}\bar\eta^{\bar z}\chi_z)(\gamma_a^i\chi_i)(\del_z\gamma_b^j\rho^z\chi_j)} 
  = \kappa \dint[_C]{dz\wedge d\bar z}\varphi_I{}_{\bar z}^z\Omega_{zij}\Tr(\gamma_a^i\del_z \gamma_b^j).
\end{equation}
There are again higher point functions with extra boundary insertions, 
following the pattern as above. 

Note that although we have derivatives of sections of $\CN_C$ we do not 
find any term involving the connection (which is part of 
the Christoffel connection $\Gamma^i_{zj}$ of $M$). However 
the total expression is still manifestly covariant. This can be seen 
by writing the relevant contributions in terms of a holomorphic 
version of the Lie derivative $\CL$,  
\begin{equation}
  (\CL_\gamma\varphi)^i = \gamma^j\del_j\varphi^i-\varphi^\mu\del_\mu \gamma^i, \qquad
  (\CL_\gamma\varphi)^\mu = \gamma^i\del_i\varphi^\mu, 
\end{equation}
which indeed is manifestly covariant --- if we would have written covariant 
derivatives the Christoffel connection cancels. 
The absence of the Christoffel connection is necessary for the model to be 
independent of the K\"ahler structure. 

The general correlation function for $\varphi_I\in H^{-1,1}(M)$ and 
with $n+1$ scalar insertions can then be written 
\begin{equation}
  \Vev{\hat\phi_I\hat\gamma_{a_0}\dint{\hat\gamma_{a_1}^{(1)}}\cdots\dint{\hat\gamma_{a_n}^{(1)}}} 
  = \kappa^n \dint[_C]{dz\wedge d\bar z}\Omega_{zij}\Tr\Bigl(
   \gamma_{a_0}^j(\CL_{\gamma_{a_1}}\cdots\CL_{\gamma_{a_n}}\varphi_I)^i_{\bar z}\Bigr).  
\end{equation}

The worldsheet field $\bar\eta^{\bi}$ has no zero-modes. Also, it 
does not give rise to any contraction. Therefore, all closed string 
operators containing this field will give vanishing correlators. 
This applies to the other components of $\varphi\in H^{-1,1}(M)$. 
Also, $\beta\in H^{0,2}(M)$ will not contribute to any correlation function. 

A similar analysis can be done for $\theta_I\in H^{-2,0}(M)$. As they do not 
contain $\bar\eta^{\bar\mu}$ we need at least one operator corresponding 
to the gauge field to soak up its zero mode. For 
$\frac{1}{2}\theta^{ij}\chi_i\chi_j$, the lowest order correlator is a 2-point 
function given by 
\begin{equation}
  \Vev{\hat\phi_I\hat\alpha_a} 
  = \frac{1}{2}\int_C\theta_I^{ij}\Omega_{zij}\Tr\Bigl(\alpha_{a\bar z}\Bigr). 
\end{equation}
For the component $\frac{1}{2}\theta^{zj}\chi_z\chi_j$ we need to contract 
the field $\chi_z$, therefore we need another descendant. This leads to a 
3-point function 
\begin{equation}
  \Vev{\hat\phi_I\hat\alpha_a\dint{\hat\gamma_b^{(1)}}} 
  = \kappa \int_C\theta_I^{zi}\Omega_{zij}\Tr(\alpha_{a\bar z}\del_z\gamma_b^j).
\end{equation}
There are higher correlators, which involve derivatives of $\theta_I$. 

The general correlation function for $\hat\phi_I$ derived from $\theta_I\in H^{-1,1}(M)$ 
and with one 1-form and $n$ scalar insertions can then be written 
\begin{equation}
  \Vev{\hat\phi_I\hat\alpha_{a_0}\dint{\hat\gamma_{a_1}^{(1)}}\cdots\dint{\hat\gamma_{a_n}^{(1)}}} 
  = \frac{\kappa^n}{2} \dint[_C]{dz\wedge d\bar z}\Omega_{zij}\Tr\Bigl(
   \alpha_{a_0\bar z}(\CL_{\gamma_{a_1}}\cdots\CL_{\gamma_{a_n}}\theta_I)^{ij}\Bigr).  
\end{equation}

\subsection{Background Gauge Fields}

We will now generalize the calculation of the mixed correlators to the presence 
of a background gauge field. We will only consider the pure gauge field case. 
The gauge field couples to the string by a boundary term 
\begin{equation}
  S_A = \dint[_{\del\Sigma}]{d\sigma} A^{(1)}
   = \dint[_{\del\Sigma}]{d\sigma}\Bigl(A_{\bar\mu}\del_\sigma{\bar z}^{\bar\mu}-iF_{\mu\bar\mu}\rho^\mu\bar\eta^{\bar\mu}\Bigr). 
\end{equation}
Equivalently, the effect of turning on a background gauge field can be accomplished 
by adding the exponentiated operator $\exp(\int\hat A^{(1)})$ in the correlators. 
This will add extra terms as there are now also contractions to the extra 
operators $\hat A^{(1)}$. 
Of course these extra contributions can also be understood in terms of 
Feynman diagrams built on vertices found above, by contracting some external 
lines to the background gauge field. For $\varphi\in H^{-1,1}(M)$ the 
extra correlators are given by 
\begin{equation}
  \Vev{\hat\varphi\hat\alpha_a} = \kappa\int\varphi\cdot\Omega\wedge\Tr(F^{1,1}\wedge\alpha_a).
\end{equation}
Similarly, for $\theta\in H^{-2,0}(M)$ there are mixed correlators 
\begin{eqnarray}
  \Vev{\hat\theta\hat\alpha_a} &=&  \frac{\kappa^2}{2}\int\theta\cdot\Omega\wedge\Tr(
   F^{1,1}\wedge F^{1,1}\wedge\alpha_a), \\
  \Vev{\hat\theta\hat\alpha_a\dint{\hat\alpha_b^{(1)}}} &=& \frac{\kappa^2}{2}\int\theta\cdot\Omega\wedge\Tr(
   \alpha_a\wedge F^{1,1}\wedge\del\alpha_b + \alpha_a\wedge\del\alpha_b\wedge F^{1,1}).
\end{eqnarray}

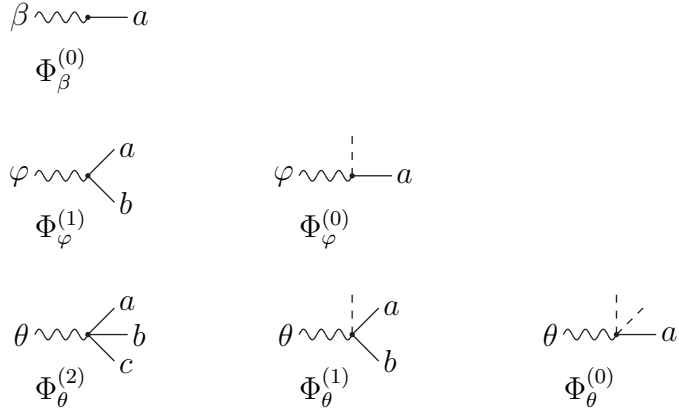
\begin{figure}[ht]
\begin{center}
\begin{picture}(245,180)
\put(0,120){%
  \begin{picture}(55,40)(10,-10)
  \Photon(20,20)(40,20)23
  \Vertex(40,20)1
  \Line(40,20)(55,20)
  \Text(18,20)[r]{$\beta$}
  \Text(57,20)[l]{$a$}
  \Text(40,0)[r]{$\Phi_\beta^{(0)}$}
  \end{picture}
}
\put(0,60){%
  \begin{picture}(50,40)(10,-10)
  \Photon(20,20)(40,20)23
  \Vertex(40,20)1
  \Line(40,20)(50,30)
  \Line(40,20)(50,10)
  \Text(18,20)[r]{$\varphi$}
  \Text(52,30)[l]{$a$}
  \Text(52,10)[l]{$b$}
  \Text(40,0)[r]{$\Phi_\varphi^{(1)}$}
  \end{picture}
}
\put(100,60){%
  \begin{picture}(55,40)(10,-10)
  \Photon(20,20)(40,20)23
  \Vertex(40,20)1
  \Line(40,20)(55,20)
  \DashLine(40,20)(40,35)3
  \Text(18,20)[r]{$\varphi$}
  \Text(57,20)[l]{$a$}
  \Text(40,0)[r]{$\Phi_\varphi^{(0)}$}
  \end{picture}
}
\put(0,0){%
  \begin{picture}(55,40)(10,-10)
  \Photon(20,20)(40,20)23
  \Vertex(40,20)1
  \Line(40,20)(50,30)
  \Line(40,20)(55,20)
  \Line(40,20)(50,10)
  \Text(18,20)[r]{$\theta$}
  \Text(52,32)[l]{$a$}
  \Text(57,20)[l]{$b$}
  \Text(52, 8)[l]{$c$}
  \Text(40,0)[r]{$\Phi_\theta^{(2)}$}
  \end{picture}
}
\put(100,0){%
  \begin{picture}(50,40)(10,-10)
  \Photon(20,20)(40,20)23
  \Vertex(40,20)1
  \Line(40,20)(50,30)
  \Line(40,20)(50,10)
  \DashLine(40,20)(40,35)3
  \Text(18,20)[r]{$\theta$}
  \Text(52,30)[l]{$a$}
  \Text(52,10)[l]{$b$}
  \Text(40,0)[r]{$\Phi_\theta^{(1)}$}
  \end{picture}
}
\put(200,0){%
  \begin{picture}(55,40)(10,-10)
  \Photon(20,20)(40,20)23
  \Vertex(40,20)1
  \Line(40,20)(55,20)
  \DashLine(40,20)(40,35)3
  \DashLine(40,20)(50,30)3
  \Text(18,20)[r]{$\theta$}
  \Text(57,20)[l]{$a$}
  \Text(40,0)[r]{$\Phi_\theta^{(0)}$}
  \end{picture}
}
\end{picture}
\end{center}
\caption{The mixed vertices of the fundamental open-closed string field theory. 
The dashed lines denote the coupling to the background gauge field. Below the diagrams 
is the corresponding map $\Phi_I^{(n)}$.}
\label{vertices}
\end{figure}

All the different vertices are summarized in \figref{vertices}. We used 
a wiggle for the closed string propagators and a line for the open string 
propagators. The coupling to the background is indicated by  a dashed line. 

A similar analysis can be performed for the scalars, by introducing a 
boundary term $S_X=\dint[_{\del\Sigma}]{d\sigma}X^{(1)}$. They can be found 
using the Feynman diagrams, introducing coupling to the background.

\subsection{Regularization and Effective Field Theory}

We now shortly discuss an important point about regularization, which will 
explain the difference between the effective and fundamental string field theory.

In calculating the correlation functions, we have to be careful about the boundaries of 
integration, where two boundary operators approach each other. We consider the 
collision of two boundary operators inside a correlation function of the form 
\begin{equation}
  \dint{d\sigma}\Vev{\cdots\hat\alpha_a^{(1)}(\sigma)\hat\alpha_b(0)\cdots}.
\end{equation}
Because we are considering a topological string theory, the contribution will 
come from the collision point itself, here the boundary of integration at $0$. 
This can be seen by introducing an auxiliary metric of magnitude $t$ on the worldsheet, 
and taking $t\to\infty$. As the theory is topological, the correlators will not depend 
on $t$. In the $t\to\infty$ limit the correlation will be naively zero, as the propagator 
is proportional to $1/t$. The only contribution can therefore 
come from the diverging collision. 
Therefore we study the contribution form this collision, and take an upper limit 
for the integral of $\lambda$. Furthermore we will also cut off the integral at 
a lower bound of $\epsilon$. We will assume that contraction of the operators 
will have a singularity in the OPE, which is of the form $\frac{1}{t\sigma}$, 
where $t$ is the magnitude of the worldsheet metric. 
As we are in a topological limit 
\cite{witcs}
\begin{equation}
  \frac{1}{t}\dint[_\epsilon^\lambda]{\frac{d\sigma}{\sigma}}\sigma^{\Delta/t} 
  = \frac{1}{\Delta}\biggl(\lambda^{\Delta/t}-\epsilon^{\Delta/t}\biggr).
\end{equation}
When we take $\epsilon\to0$ first, then the $t\to\infty$ limit would produce an 
insertion of the propagator $\frac{1}{\Delta}$. However, when we adopt a 
point splitting procedure by taking $\epsilon\to0$ only at the end of the calculation, 
we first take the $t\to\infty$ limit, and this will give a vanishing contribution. 

In the calculations above we used the point splitting regularization, 
$\epsilon>0$. Let us now see what we get when we take the alternative 
regularization with $\epsilon=0$. To discern from the correlators in the 
point splitting regularization we will denote the correlators and maps by 
$\tilde\Phi_{Ia_0\ldots a_n}$ and $\tilde\Phi_I^{(n)}$, respectively. 

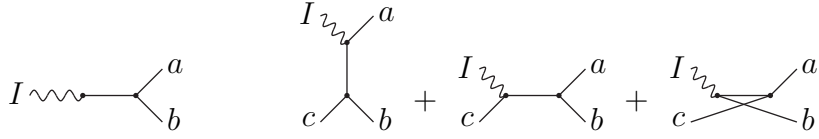
\begin{figure}
\begin{center}
\begin{picture}(310,70)
\put(0,0){%
  \begin{picture}(70,30)(0,5)
  \Photon(10,20)(30,20)23
  \Line(30,20)(50,20)
  \Line(50,20)(60,30)
  \Line(50,20)(60,10)
  \Vertex(30,20)1
  \Vertex(50,20)1
  \Text( 8,21)[r]{$I$}
  \Text(62,31)[l]{$a$}
  \Text(62,11)[l]{$b$}
  \end{picture}
}
\put(110,0){%
  \begin{picture}(40,50)(0,5)
  \Photon(10,50)(20,40)23
  \Line(10,10)(20,20)
  \Line(30,10)(20,20)
  \Line(30,50)(20,40)
  \Line(20,20)(20,40)
  \Vertex(20,20)1
  \Vertex(20,40)1
  \Text( 8,51)[r]{$I$}
  \Text(32,51)[l]{$a$}
  \Text(32,11)[l]{$b$}
  \Text( 8,11)[r]{$c$}
  \end{picture}
}
\put(155,10){$+$}
\put(170,0){%
  \begin{picture}(60,30)(0,5)
  \Photon(10,30)(20,20)23
  \Line(10,10)(20,20)
  \Line(50,10)(40,20)
  \Line(50,30)(40,20)
  \Line(20,20)(40,20)
  \Vertex(20,20)1
  \Vertex(40,20)1
  \Text( 8,31)[r]{$I$}
  \Text(52,31)[l]{$a$}
  \Text(52,11)[l]{$b$}
  \Text( 8,11)[r]{$c$}
  \end{picture}
}
\put(235,10){$+$}
\put(250,0){%
  \begin{picture}(60,30)(0,5)
  \Photon(10,30)(20,20)23
  \Line(10,10)(40,20)
  \Line(50,10)(20,20)
  \Line(50,30)(40,20)
  \Line(20,20)(40,20)
  \Vertex(20,20)1
  \Vertex(40,20)1
  \Text( 8,31)[r]{$I$}
  \Text(52,31)[l]{$a$}
  \Text(52,11)[l]{$b$}
  \Text( 8,11)[r]{$c$}
  \end{picture}
}
\end{picture}
\end{center}
\caption{The Feynman diagram contributing to the effective mixed 3-point 
function $\tilde\Phi_{Iab}$ and the three Feynman diagrams contributing to the 
mixed 4-point function $\tilde\Phi_{Iabc}$.}
\label{feynman4pt}
\end{figure}

Let us calculate the extra contributions to the mixed 3-point function
\begin{equation}
  \tilde\Phi_{Iab}  = \Vev{\hat\phi_I\hat\alpha_a\dint{\hat\alpha_b^{(1)}}},
\end{equation}
coming from the collision of boundary operators. 
First, we have to insert the first descendants. We get 
\begin{equation}
\tilde\Phi_{Iab}  = \Phi_{Iab} + \dint{d\sigma} 
\Bigl\langle (\varphi_I\bar\eta\chi)(u)(\alpha_a\bar\eta)(0)
  (\alpha_b\del_\sigma{\bar z}+\del\alpha_b\rho\bar\eta)(\sigma)\Bigr\rangle.
\end{equation}
There are obvious contractions of indices, which we only indicated by the brackets. 
The contraction of $\rho$ with the closed string operator was responsible 
for the correlation function \eqref{def11}; here we are interested in the 
other contributions. Using that we can calculate this in the limit $t\to\infty$, 
these can only come from the poles in the integral where 
vertex operators collide, at the boundaries of the integration. There are 
$z\del_\sigma\bar z$ and $\bar\eta\rho$ contractions, which can be summarized as 
\begin{equation}
\tilde\Phi_{Iab}  = \Phi_{Iab} + \frac{1}{t}\dint{d\sigma} \frac{1}{\sigma}
\Bigl\langle (\varphi_I\bar\eta\chi)(u)\sigma^{\Delta/t}\bar\eta\bar\del^\dagger(\alpha_a\alpha_b)(0)\Bigr\rangle,
\end{equation}
The field $\chi_\mu$ in the closed string operator, having no zero-modes, 
should contract to a background gauge field. This also provides the extra 
$\bar\eta$ needed to soak up the third zero-mode. 
Replacing the correlators with zero-mode integrals, we find 
in the $t\to\infty$ limit 
\begin{equation}
\tilde\Phi_{Iab}  = \Phi_{Iab} + \kappa\int\varphi_I\cdot\Omega\wedge\Tr\biggl( 
F^{1,1}\wedge\frac{\bdA^\dagger}{\Delta_A}\Bigl(\alpha_a\wedge\alpha_b\Bigr)\Biggr).
\end{equation}
This expression has a nice interpretation in terms of Feynman diagrams. 
Noting that $\frac{\bdA^\dagger}{\Delta_A}$ can be identified with 
the propagator $\bar\del\inv$ (after a gauge fixing), this expression 
can be identified with the first Feynman diagram depicted in \figref{feynman4pt}. 
Here we used the vertices derived earlier in a background gauge field, 
not explicitly depicting the coupling to the background. 

The mixed four-point function 
\begin{equation}
  \tilde\Phi_{Iabc}=\Vev{\hat\phi_I\hat\alpha_a\dint{\hat\alpha_b^{(1)}\dint{\hat\alpha_c^{(1)}}}}
\end{equation}
can be calculated in a similar way. Now because there are two integrations, 
we find two contractions; between every adjacent pair. Remember that the 
boundary of a disc is a circle, so there is also a collision between 
$\alpha_a$ and $\alpha_b$. The result is 
\begin{eqnarray}
  \tilde\Phi_{Iabc} = \kappa\int\varphi_I\cdot\Omega\wedge\Tr\biggl(
  \del\alpha_a\wedge\frac{\bdA^\dagger}{\Delta_A}(\alpha_b\wedge\alpha_c)
  \biggr) \pm\mbox{2 perms}.
\end{eqnarray}
The three cyclic permutations in this formula can be interpreted in 
terms of the three tree level Feynman diagrams for this correlator, 
as depicted in Figure \ref{feynman4pt}.

In general, the collision between adjacent operators generate all 
tree level Feynman diagrams. We conclude that this regularization 
produces the effective field theory. 
In the background gauge field the BRST operator will also change to the covariant 
operator $\bdA$. This can of course be understood as usual by a shift of the 
open string field. 

In the effective theory, there are corrections from collisions between 
open string operators and the background gauge field. The propagator 
$\bdA\inv=\frac{\bar\del^\dagger}{\Delta}$ gets contributions from a 
Dyson series involving tadpoles to the background. This leads to the covariant 
form of the propagator $\bdA\inv=\frac{\bdA^\dagger}{\Delta_A}$, as depicted in 
\figref{effprop}. 

\begin{figure}[ht]
\begin{center}
\begin{picture}(200,70)(0,-10)
\Line(0,19)(50,19)
\Line(0,21)(50,21)
\Text(25,0)[m]{$\frac{\bdA^\dagger}{\Delta_A}$}
\Text(70,20)[m]{$=$}
\Text(100,20)[m]{$\displaystyle\sum_n$}
\Line(120,20)(200,20)
\DashLine(140,20)(140,40)3
\DashLine(150,20)(150,40)3
\DashLine(160,20)(160,40)3
\DashLine(180,20)(180,40)3
\Vertex(140,20)1
\Vertex(150,20)1
\Vertex(160,20)1
\Vertex(180,20)1
\Text(170,30)[m]{$\cdots$}
\Text(160,45)[b]{$\stackrel{n\times}{\overbrace{\rule{40pt}{0pt}}}$}
\Text(160,0)[m]{$\Bigl(\frac{\bar\del^\dagger}{\Delta}A\Bigr)^n\frac{\bar\del^\dagger}{\Delta}$}
\end{picture}
\end{center}
\caption{Dyson series for the propagator in a background gauge field, 
leading to the covariant propagator.}
\label{effprop}
\end{figure}

The effective theory is relevant for the superpotential of the dimensionally reduced 
theory. This has been studied recently in the context of the 
open string B-model in \cite{laza,tomas,dogojato}.

\section{Closed Strings as Deformations}
\label{sec:def}

\subsection{Deformations}

A complex structure deformation $\varphi\in H^{-1,1}(M)$ acts on the covariant 
antiholomorphic derivative as 
\begin{equation}
\delta\bdA = \kappa\varphi\cdot\del, 
\end{equation}
as can be seen from \eqref{def11}. 
This should be compared to the formula for the variation of the 
BRST operator in closed string theory due to a deformation by an 
operator $\hat\phi$, which is given by \cite{verl} 
\begin{equation}
\delta Q = \oint\hat\phi^{(1)}. 
\end{equation}
We observe that the operator $\kappa\varphi\cdot\del$ has the interpretation of 
the integral of the action of the first descendant of the corresponding 
bulk operator.

Next, we consider the deformations \eqref{def20}. They correspond to a 
deformation of the cubic term in the effective action, and therefore 
a deformation of the product. Indeed, the first order deformation by an 
element $\theta\in H^{-2,0}(M)$ can be written 
\begin{equation}
  \frac{\kappa^2}{2}\int\Omega\wedge\Tr(
   \alpha_a\wedge\theta^{ij}\del_i\alpha_b\wedge\del_j\alpha_c). 
\end{equation}
This is precisely the first order deformation for the noncommutative 
star-product corresponding to the bivector $\theta$, 
\begin{equation}
  \alpha_a\star_\theta\alpha_b = 
  \alpha_a\wedge\alpha_b+\frac{\kappa^2}{2}\theta^{ij}\del_i\alpha_a\wedge\del_j\alpha_b
   +\CO(\kappa^4\theta^2)
\end{equation}
This was expected, as this deformation corresponds (among others) to a $B$-field, 
which we know induces this star-product. In fact, we will argue later that 
higher order correlators in $\theta$ will generate the full star product, given 
by Kontsevich's formula for deformation quantization. 

The last deformation \eqref{def02} corresponds to a shift of the field-strength 
$F^{0,2}$ by the corresponding element of $H^{0,2}(M)$. Note that here and in 
the deformation above we probably used a different regularization than usually 
in the context of noncommutative gauge theories, as here there is an explicit 
shift of the field strength, as in the case of the commutative description, 
while in the former case the full $(-2,0)$-form contributes to the star-product, 
and not only the real part (corresponding to the $B$-field; the imaginary part 
corresponds to the K\"ahler form).

\subsection{Descent Equations}

In calculating the correlation functions, we saw that apart from the 
leading correlation functions, there were also correlation functions 
with less open string insertions. In these correlation functions, the 
$(1,1)$-part of the field strength $F^{1,1}$ plays an important role. 
Here we explain the interpretation of these lower order correlation 
functions and how $F^{1,1}$ comes in. 

To understand it somewhat better, we interpret the correlation functions 
again in terms of multilinear maps on the open string algebra, as 
in \eqref{corrhoch}. We start from the correlation functions involving 
a closed string operator associated with $\varphi\in H^{-1,1}(M)$, related to 
a complex structure deformation. For such an operator, the leading 
component with the highest number of open string operators was given in 
\eqref{def11} of open string operators had two boundary operator. 
Therefore it corresponds to a linear map, and write this correlator 
as $\Vev{\hat\alpha_a\Phi_\varphi^{(1)}(\hat\alpha_b)}$. 
The next highest order map represented by the mixed 2-point function 
will be denoted as $\Vev{\hat\alpha_a\Phi_\varphi^{(0)}}$. 
We see from the explicit formulas of the correlation functions that 
\begin{equation}
  \Phi_\varphi^{(1)}(\alpha)=\varphi\cdot\del\alpha,\qquad 
  \Phi_\varphi^{(0)}=\varphi\cdot F^{1,1}.
\end{equation}
To understand the relation between these components, and especially the 
way the field strength arises, consider the following identity, 
\begin{equation}\label{descent}
\{\bdA,\varphi\cdot\del\} = -\varphi\cdot F^{1,1}. 
\end{equation}
This equation should be read as an equation for operators acting 
on adjoint forms $\alpha\in \Omega^{0,*}_{\bdA}(M,\End(M))$,
\begin{equation}
 \bdA(\varphi\cdot\del\alpha) + \varphi\cdot\del(\bdA\alpha) 
 = -(\varphi\cdot F^{1,1})\wedge\alpha + \alpha\wedge(\varphi\cdot F^{1,1}). 
\end{equation}

This relation is directly related to the descent 
equation in the closed string, which reads $\{Q,\phi^{(1)}\}=d\phi$. 
In the open string B-model, the operator $\bdA$ was the BRST operator. 
Notice that integrating $d\phi$ on a half circle around a boundary operator 
can be written, using Stokes, as two boundary terms, of $\phi$ moved 
to the boundary to either side of the boundary operator. This then 
corresponds to a commutator with the boundary operator induced by $\phi$,
see \figref{bndyact}. On the other hand, integrating $\phi^{(1)}$ over a 
half-circle produces the action $\Phi^{(1)}$ on the boundary operator. 
From this, we learn the effect of a boundary operator induced by a 
bulk operator: the bulk operator corresponding to $\varphi$ acts by 
commutation with $\varphi\cdot F^{1,1}$. In fact, we can only read off the action 
by a commutator, and not the (star) product. The latter one can of course 
easily be guessed to be just the corresponding action with just the 
wedge product. For most applications, this will not be needed however. 

\begin{figure}
\begin{center}
\begin{picture}(320,60)(0,0)
\put(0,0){%
  \begin{picture}(80,60)(0,-15)
    \Line(0,0)(80,0)
    \Vertex(40,0)2
    \Vertex(20,0)1
    \Vertex(60,0)1
    \ArrowArc(40,0)(20,0,180)
    \Text(40,35)[c]{$\doint{Q\phi^{(1)}}=\doint{d\phi}$}
    \Text(40,-10)[c]{$\alpha$}
  \end{picture}} 
\put(95,15){$=$}
\put(120,0){%
  \begin{picture}(80,35)(0,-15)
    \Line(0,0)(80,0)
    \Vertex(40,0)2
    \Vertex(20,0)2
    \Text(20,10)[c]{$\phi$}
    \Text(40,-10)[c]{$\alpha$}
  \end{picture}}
\put(215,15){$-$}
\put(240,0){%
  \begin{picture}(80,70)(0,-15)
    \Line(0,0)(80,0)
    \Vertex(40,0)2
    \Vertex(60,0)2
    \Text(60,10)[c]{$\phi$}
    \Text(40,-10)[c]{$\alpha$}
  \end{picture}}
\end{picture}
\end{center}
\caption{The relation between the descent equation in the bulk, and the 
action by a commutator at the boundary.}
\label{bndyact}
\end{figure}

This relation also reflects the descent equation in Hochschild cohomology, 
which were derived in \cite{homa} from a Ward identity. 
The anticommutator with the BRST operator $\{Q,\Phi_\varphi^{(1)}\}$ in 
the left-hand side of \eqref{descent} is precisely the action of 
the coboundary $\delta_Q$ in the Hochschild double complex. The right-hand 
side can actually be interpreted in terms of the usual Hochschild coboundary, 
related to the product, $\delta_m\Phi_\varphi^{(0)}$. Hence we can write 
this equation in the form 
\begin{equation}
\delta_m\Phi_\varphi^{(0)}=-\delta_Q\Phi_\varphi^{(1)}.
\end{equation}
This is the descent equation in the Hochschild double complex with 
the two coboundaries $\delta_Q$ and $\delta_m$. 
Notice that there is no lower descendant, as $\delta_Q\Phi_\varphi^{(0)}=0$ 
due to the Bianchi identity and $F^{0,2}=\bdA^2=0$. Also note that the 
top component satisfies $\delta_m\Phi_\varphi^{(1)}=0$, which is the 
derivation condition of $\varphi\cdot\del$. 

The relation between the descent equation can also be performed for the 
other closed string operators. For $\theta\in H^{-2,0}(M)$ we can write 
equation \eqref{def20} in terms of a map 
$\Phi_\theta^{(2)}$ as $\Vev{\hat\alpha_a\Phi_\theta^{(2)}(\hat\alpha_b,\hat\alpha_c)}$. 
The following equation we write as $\Vev{\hat\alpha_a\Phi_\theta^{(1)}(\hat\alpha_b)}$, 
and the next as $\Vev{\hat\alpha_a\Phi_\theta^{(0)}}$. From the explicit 
expressions for the correlators we have 
\begin{equation}
\Phi_\theta^{(2)}(\alpha_a,\alpha_b)
 = \frac{1}{2}\theta\cdot(\del\alpha_a\wedge\del\alpha_b),\quad
\Phi_\theta^{(1)}(\alpha_a) = \theta\cdot(F^{1,1}\wedge\del\alpha_a),\quad
\Phi_\theta^{(0)} = \frac{1}{2}\theta\cdot(F^{1,1}\wedge F^{1,1}).
\end{equation}
Using the above relation, one easily sees that the latter two maps are 
indeed descendants, that is $\delta_m\Phi_\theta^{(1)}=-\delta_Q\Phi_\theta$ 
and $\delta_m\Phi_\theta^{(2)}=-\delta_Q\Phi_\theta^{(1)}$. Also, indeed 
$\Phi_\theta$ can be considered a deformation of the wedge-product. 

For the element $\beta\in H^{0,2}(M)$ we write \eqref{def02} as 
$\Vev{\hat\alpha_a\Phi_\beta}$. That is, we simply have $\Phi_\beta=\beta$. 
There are no other descendants, as $\delta_Q\Phi_\beta=\bar\del\beta=0$ 
and $\delta_m\Phi_\beta=[\beta,\cdot]=0$.

\subsection{Relation to the Hochschild Complex}

The closed string operators correspond to multilinear maps on the open string 
algebra $\CA=\Omega^{0,*}(M,\End(E))$, which satisfy descend equation for 
on-shell closed string algebras as we saw above. This just says that they 
correspond to elements of the Hochschild cohomology. Therefore, we find a 
direct correspondence between the closed string BRST cohomology 
which was identified with $H^{-*,*}(M)$ and the Hochschild cohomology 
$H(\CA)$. Using this, we can actually identify them, as we will see in 
more detail later. On the other hand, the correlation functions give explicit 
elements of the Hochschild complex $C^*(\CA,\CA)$ of multilinear maps. This 
can be seen as a formality map for the corresponding Hochschild complex. 
Here, we will give the clarify this relation by relating the element 
in cohomology with the leading component in the Hochschild cohomology. 

The number of elements in the closed string cohomology $H^{-*,*}(M)$ 
can be summarized in terms of the Hodge diamond of the Calabi-Yau manifold. 
This Hodge diamond, in the slightly unusual notation of the more natural 
cohomology, is given by 
\begin{equation}
\begin{array}{c}
\hodge{-3,0} \\
\hodge{-2,0}\hodge{-3,1} \\
\hodge{-1,0}\hodge{-2,1}\hodge{-3,2} \\
\hodge{ 0,0}\hodge{-1,1}\hodge{-2,2}\hodge{-3,3} \\
\hodge{ 0,1}\hodge{-1,2}\hodge{-2,3} \\
\hodge{ 0,2}\hodge{-1,3} \\
\hodge{ 0,3}
\end{array}
=
\begin{array}{c}
\hodge{0,0} \\
\hodge{1,0}\hodge{0,1} \\
\hodge{2,0}\hodge{1,1}\hodge{0,2} \\
\hodge{3,0}\hodge{2,1}\hodge{1,2}\hodge{0,3} \\
\hodge{3,1}\hodge{2,2}\hodge{1,3} \\
\hodge{3,2}\hodge{2,3} \\
\hodge{3,3}
\end{array}
\end{equation}
where $h^{-p,q}=\dim H^{-p,q}(M)=h^{3-p,q}$. 

Any element $\phi_I$ of the closed string cohomology corresponds through 
the correlation functions to a sequence of  multilinear maps $\Phi_I^{(n)}$. 
These maps correspond to elements of the Hochschild cohomology of the open 
string algebra. As we saw, the sequence of maps satisfy the descent equations 
$\delta_m\Phi_I^{(n)}=-\delta_Q\Phi_I^{(n+1)}$. We assume that these will terminate 
at a certain maximum value of $n$ for any given $\phi_I$. Indeed, this was 
true in the cases we studied. We will concentrate on this leading component 
with the largest order. Due to the relation between the closed string 
cohomology, the Dolbeault cohomology, and the Hochschild cohomology 
there should be a natural correspondence between the Hodge diamond 
and the Hochschild cohomology. To motivate this relation, let us look 
at what we found. For $\phi_I\in H^{-p,q}(M)$ with $p+q=2$ we already 
saw the relation between the element in cohomology and the leading order map. 
They are summarized in \tbref{hochcorr}. 
\begin{table}[ht]
\[
\renewcommand{\arraystretch}{1.5}
\begin{array}{|c|c|ccc|}
\hline 
(-p,q) & \hat\phi_I & \Phi_I^{(0)} & \Phi_I^{(1)} & \Phi_I^{(2)} \\ \hline
(-2,0) & \frac{1}{2}\theta^{\mu\nu}\chi_\mu\chi_\nu & 
 \frac{1}{2}\theta^{\mu\nu}F_{\mu\bar\mu}F_{\nu\bar\nu}\bar\eta^{\bar\mu}\bar\eta^{\bar\nu} & 
 \theta^{\mu\nu}F_{\mu\bar\mu}\bar\eta^{\bar\nu}\del_\nu & \frac{1}{2}\theta^{\mu\nu}\del_\mu\wedge\del_\nu \\
(-1,1) & \varphi^\mu_{\bar\mu}\bar\eta^{\bar\mu}\chi_\mu & 
 \varphi^\mu_{\bar\mu}F_{\mu\bar\nu}\bar\eta^{\bar\nu}\bar\eta^{\bar\nu} & \varphi^\mu_{\bar\mu}\bar\eta^{\bar\mu}\del_\mu & \\
(0,2)  & \frac{1}{2}\beta_{\bar\mu\bar\nu}\bar\eta^{\bar\mu}\bar\eta^{\bar\nu} & 
 \frac{1}{2}\beta_{\bar\mu\bar\nu}\bar\eta^{\bar\mu}\bar\eta^{\bar\nu} && \\ \hline
\end{array}
\]
\caption{Correspondence of elements in the cohomology to elements 
of the Hochschild complex, for $p+q=2$.}
\label{hochcorr}
\end{table}
These correspondences suggest 
that the leading component for an element $\phi_I\in H^{-p,q}(M)$ 
is a map $\Phi_I^{(p)}\in \Hom(\CA^{\otimes p},\CA)^q$ of order $p$ and 
degree $q$. We will argue below that this is in fact true. 
This leads us to organize the Hochschild complex in accordance 
to the Hodge diamond, as depicted in \tbref{hoch}. 
\begin{table}[h]
\renewcommand{\arraystretch}{1.5}
\begin{center}
\begin{tabular}{c}
\hochbox{3}{0} \\
\hochbox{2}{0}\hochbox{3}{1} \\
\hochbox{1}{0}\hochbox{2}{1}\hochbox{3}{2} \\
\hochbox{0}{0}\hochbox{1}{1}\hochbox{2}{2}\hochbox{3}{3} \\
\hochbox{0}{1}\hochbox{1}{2}\hochbox{2}{3} \\
\hochbox{0}{2}\hochbox{1}{3} \\
\hochbox{0}{3}
\end{tabular}
\end{center}
\caption{The relevant portion of the Hochschild complex, organized according 
to the Hodge diamond.}
\label{hoch}
\end{table}

Let us motivate this more general relation. Remember that closed string 
operators in the BRST cohomology where functions of 
$z^\mu,\bar z^{\bar\mu},\bar\eta^{\bar\mu},\chi_\mu$. 
The degree $p$ corresponds to the number of $\bar\eta$'s, and the degree $q$ 
to the number of $\chi$'s. To find the leading component of the map in the 
Hochschild complex, one replaces all the $\chi_\mu$ by the differential 
operator $\del_\mu$. If we start from a $(-p,q)$-form $\phi_I$, defined as in 
\eqref{clop}, this gives a $p$-linear map $\Phi_I^{(p)}$ of degree $q$ 
acting on the open string algebra $\CA$. This map is given by the degree 
$q$ polydifferential operator 
\begin{equation}
\Phi_I^{(p)} = 
  \phi^{\mu_1\ldots \mu_p}_{\bar\mu_1\ldots\bar\mu_q}\bar\eta^{\bar\mu_1}\cdots\bar\eta^{\bar\mu_q}
  \del_{\mu_1}\wedge \cdots\wedge\del_{\mu_p}. 
\end{equation}
This is indeed an element of $\Hom(\CA^{\otimes p},\CA)^q$, confirming 
the above placement in the Hodge diamond. It can also be seen that this will correspond 
to the correlation function, as can be seen by fully contracting all $\chi_\mu$ 
to $z^\mu$ in different boundary operators, therefore acting indeed as $\del_\mu$. 
These maps, for $\phi_I$ an element of the cohomology, can indeed be seen to be 
closed with respect to the coboundary $\delta_m$. Again, there are descendants 
$\Phi_I^{(n)}$ for $0\leq n<p$, which can be found by replacing some $\chi_\mu$ 
by $F^{1,1}$ 
\begin{equation}
\Phi_I^{(n)} = \kappa^p
  \phi^{\mu_1\ldots\mu_p}_{\bar\mu_1\ldots\bar\mu_q}\bar\eta^{\bar\mu_1}\cdots\bar\eta^{\bar\mu_q}
  \bar\eta^{\bar\nu_1}\cdots\bar\eta^{\bar\nu_{p-n}}F_{\mu_1\bar\nu_1}\cdots F_{\mu_{p-n}\bar\nu_{p-n}}
  \del_{\mu_{p-n+1}}\wedge \cdots\wedge\del_{\mu_p} + perms. 
\end{equation}
These have degrees such that the total degree in the Hochschild 
complex is constant, equal to $p+q$. They also satisfy the descent equations 
$\delta_m\Phi_I^{(n)}=-\delta_Q\Phi_I^{(n+1)}$. In particular, the leading 
component of order $p$ is closed with respect to $\delta_m$. In general, 
we can summarize what we said above by replacing the fermion $\chi_\mu$ 
with a ``covariant'' holomorphic derivative 
\begin{equation}
  \chi_\mu \to \kappa\frac{D}{Dz^\mu} 
  = \kappa\frac{\del}{\del z^\mu}+\kappa F_{\mu\bar\mu}\bar\eta^{\bar\mu}+\kappa\del_\mu X^i\chi_i,
\end{equation}
which is closed with respect to the total coboundary $\delta_m+\delta_Q$. 
We also included a term involving the scalar for completeness.

\subsection{Observables and an Open-Closed SFT Action}

From the above correlation functions we can write down the gauge invariant 
observable (first order) $\CO_\Phi$ for $\phi\in\bigoplus_p H^{-p,q}(M)$ 
for $p+q=2$ in the following form 
\begin{equation}
\CO_\Phi = \int\phi\cdot\Omega\wedge\Tr\biggl(A+\frac{\kappa}{2}A\wedge\del A+\frac{\kappa^2}{6}A\wedge\del A\wedge\del A\biggr) 
  = \int\phi\cdot\Omega\wedge\Tr\biggl(f(\kappa F^{1,1})\wedge A\biggr),
\end{equation}
where in the last expression we introduced the function 
$f(x)=\frac{\e^x-1}{x}=1+\frac12x+\frac16x^2+\cdots$, and $f(\kappa F^{1,1})$ 
should be understood in terms of a Taylor expansion, using the wedge product. 
This function is such that $\del\Bigl(f(\kappa F^{1,1})\wedge\kappa A\Bigr)=\e^{\kappa F^{1,1}}-1$. 

As explained before, this observable can be understood as a first order deformation 
of the open string field theory action $S_0$. In the context of an open-closed string 
field theory for the B-model, we can interpret $\phi$ as the (physical) on-shell 
closed string field. The deformed action $S=S_0+\CO_{\bar\Phi}$ can then be interpreted 
as a first approximation to the action of the open-closed string field theory, 
with the closed string taken on-shell. 

By inserting the different components, we recover the amplitudes calculated above. 
Indeed we can write the above as (at least when we expand $\star_\theta$ to first order 
in $\theta$)
\begin{equation}
S = \int\Omega\wedge\Tr\Bigl( 
  \beta\wedge A + \frac{1}{2}A\star_\theta(\bar\del+\kappa\varphi\cdot\del)A +\frac{1}{3}A\star_\theta A\star_\theta A\Bigr) + \CO(\phi^2),
\end{equation}
with $\beta\in H^{0,2}(M)$, $\varphi\in H^{-1,1}(M)$, and 
$\theta=\kappa\inv\delta\star\in H^{-2,0}(M)$,  $\phi=\beta+\varphi+\theta$. Or in other words, 
we have the deformed action with deformed data $Q_\varphi=\bar\del+\kappa\varphi\cdot\del$, 
$\star_\theta=\wedge+\kappa^2\theta\cdot\del\wedge\del+\CO(\kappa^4\theta^2)$, and 
the same integral. Only, there is a tadpole proportional to $\beta$. 

We should remark that the full action is still cubic in $A$, although it does not 
look like it from the above form. The form degrees indeed forbid more than three 
appearances of $A$ in the action. This is very nontrivial, and only works 
because we are working in 3 complex dimensions. 

For general on-shell bulk operators $\phi\in H^{-p,q}(M)$ with $p+q\leq2$ 
we expect for the total action 
\begin{equation}
\CO_\Phi = \int\phi\cdot\Omega\wedge\Tr\biggl(f(\kappa F^{1,1})\wedge
\Bigl(A +\bar\del A +\frac{1}{2}A\wedge A 
  +\frac{1}{2}A\wedge\bar\del A +\frac{1}{3}A\wedge A\wedge A\Bigr)\biggr).
\end{equation}
For $p+q>2$ we should probably also consider non-physical open string operators. 
This is to be expected, as these operators have too high a ghost number 

Let us now also include the scalars into our discussion. 
The two interactions discussed earlier give rise to the following two terms in the 
action of the open-closed string field theory 
\begin{equation}
  \sum_n\frac{\kappa^{n+1}}{(n+1)!}\int_C\Omega_{\mu ij}\STr\Bigl((X^k\del_k)^n\varphi^i_{\bar\mu}X^j\Bigr) 
 + \sum_n\frac{\kappa^{n+2}}{(n+2)!} \int_C\Omega_{\nu ij}\STr\Bigl((X^k\del_k)^n\varphi^\nu_{\bar\mu}X^i\del_\mu X^j\Bigr),
\end{equation}
where $\STr$ denotes the symmetrized trace. 
Note that the contribution of $\e^{\kappa X^k\del_k}\varphi$ can be 
understood as a Taylor expansion, replacing the dependence of the 
normal coordinate $z^i$ in $\varphi$ by $\kappa X^i$. So indeed, 
we can understand the scalar fields as describing movement in the 
normal direction. 

A similar analysis can be done for $\theta\in H^{-2,0}(M)$. 
The correlators give the following two terms in the action 
\begin{equation}
  \frac{1}{2}\int_C\theta^{ij}\Omega_{zij}\Tr\Bigl(A_{\bar z}\Bigr)
  + \kappa \int_C\theta^{zi}\Omega_{zij}\Tr(A_{\bar z}\del_zX^j).
\end{equation}

Ignoring the transverse derivatives of $\phi$ for the moment, 
we can summarize the first order open-closed string action 
in the form 
\begin{equation}\label{prelim}
  \CO_\Phi = \int\phi\cdot\Omega\cdot\Tr\biggl(f(\kappa F^{1,1}+\kappa\del X)\wedge(A+X)\biggr),
\end{equation}
where $\cdot$ denotes the contraction of any holomorphic vector 
index with a holomorphic form index, and $\wedge$ denotes wedge 
products both for form and vector indices. 

To include the transverse derivatives, we note that there are two 
types of terms involving scalars $X$ and derivatives: 
terms involving $X\cdot\del\phi$ and of the form $\phi\cdot\del X$. 
In fact, covariance forces them to appear in a particular combination, 
which we can interpret a the holomorphic Lie-derivative $\CL_X$. 
This is most conveniently written in terms of action on forms. 
For this action we can write  
\begin{equation}\label{lie}
  \CL_X=i_X\del+\del i_X, 
\end{equation}
giving indeed the two types of terms we found. A covariant form 
involving the transverse derivatives can be obtained by replacing 
$\del X$ in \eqref{prelim} with $\CL_X$, where the derivatives 
are supposed to act on $\phi\cdot\Omega$.   

We can give a convenient covariant description, which also shows its 
form as a generalized Chern-Simons term. For this, we consider 
a family of Calabi-Yau manifold $M\times\C$. The brane $C$ is 
replaced by a family $C_y$ parametrized by $y\in\C$, which 
reduces to $C_0=C$ at the origin. 
On this brane the open string fields are allowed to have an 
auxiliary dependence on the transverse coordinates $z^i$. 
At $y=0$ the boundary condition reduces to $z^i=0$. 

Above we have noticed the role of the gauge field and the scalars. 
The effect of a background gauge field on the correlators 
can be summarized by the insertion of $\e^{\kappa F^{1,1}}$. 
The effect of a background scalar is to translate the brane, 
by the action of the Lie-derivative $\CL_X$.  
We now combine the two effects into a single exponential. 
Using the fact that the closed string field $\phi$ is on shell, 
this exponential is actually closed, that is we have a descent-like equation 
\begin{equation}
  \Tr\Bigl(\e^{\kappa F^{1,1}+\kappa\CL_X}(\phi\cdot\Omega)\Bigr) 
  = \kappa\del Y(\phi;A,X) + \Tr(\phi\cdot\Omega),
\end{equation}
for some Chern-Simons form $Y$. We could of course express $Y$ in 
terms of the function $f$ as above. To show this we use \eqref{lie}, 
and observe that the first term will never contribute, as all 
terms are $\del$-closed. We can then write the observable as 
\begin{equation}
  \CO_\Phi = \int_CY = \int_D\Tr\Bigl(\e^{\kappa F^{1,1}+\kappa\CL_X}(\phi\cdot\Omega)\Bigr) 
  - \int_D\Tr(\phi\cdot\Omega), 
\end{equation}
In the last expression we used an auxiliary chain $D\subset M\times\C$ 
such that $\del D=C$. Note that when $C=M$ this gives the correct 
expression, as we remarked before. Also when there are no gauge 
fields in the expression (this happens when $q=\dim_\C C$), the formula 
can be interpreted as a translation of the brane. Writing $\omega=\phi\cdot\Omega$ 
the integrand can be written 
\begin{equation}
  \sum_n\frac{\kappa^n}{(n+1)!}i_X(\del i_X)^n\omega
   = i_X\omega + \frac{\kappa}{2!}i_X\del (i_X\omega)+\frac{\kappa^2}{3!}i_X\del (i_X\del (i_X\omega))+\cdots.
\end{equation}
which appeared in a special situation in \cite{kklm1,kklm2}.

\section{The Hochschild Cohomology of HCS}
\label{sec:hoch}

In this section we explicitly calculate the Hochschild cohomology of the open string algebra 
of the B-model. To do the calculation, we will use Hochschild-Kostant-Rosenberg 
theorem reviewed in \appref{poly}.

\subsection{Calculation of the Hochschild Cohomology}

We take for the open string algebra the full off-shell algebra 
$\CA=\Omega^{0,*}(M,\End(E))$ of endomorphism valued $(0,p)$-forms. This 
has the structure of a differential graded associative algebra (dg-algebra), 
with differential $\bdA$ and the product is the combination of the wedge product 
and the local matrix product in $\End(E)$. 

To see what happens we will first take a look at the case where $E$ is the trivial 
$\U(1)$ bundle, i.e.\ we take $\CA=\Omega^{0,*}(M)$. Later we shall argue that 
we can reduce to this case also in the situation of a nontrivial bundle, 
making use of the Morita equivalence between these algebras. 

We first look at the most trivial case of a flat CY manifold $M=\C^3$ with a trivial 
bundle, so that $\bdA=\bar\del$. We can reduce the problem to the much 
better behaved problem of Hochschild cohomology of a polynomial algebra. Note that 
the polynomial forms are dense in the algebra. So we replace the algebra $\CA$ by 
the polynomial algebra $\CA=\C[z^\mu,\bar z^{\bar\mu},\bar\eta^{\bar\mu}]$, where $z^\mu$ and 
$\bar z^{\bar\mu}$ are generators of degree 0, and $\bar\eta^{\bar\mu}$ is a Grassmann generator 
if degree 1. The differential on this algebra can be written 
$Q=\bar\eta^{\bar\mu}\frac{\del}{\del\bar z^{\bar\mu}}$. If this differential were zero, 
the Hochschild cohomology would be the cohomology of the multilinear maps with respect 
to the Hochschild coboundary $\delta_m$, $\HH^*(\CA)=H^*_{\delta_m}(C^*(\CA,\CA))$.
This is precisely the situation handled by the Hochschild-Kostant-Rosenberg theorem, 
giving the polynomial algebra 
\begin{equation}
\CC = H^*_{\delta_m}(C^*(\CA,\CA))
 = \C[z^\mu,\bar z^{\bar\mu},\bar\eta^{\bar\mu};\chi_\mu,\bar\chi_{\bar\mu},\bar p_{\bar\mu}],
\end{equation}
where the extra generators $\chi_\mu$ and $\bar\chi_{\bar\mu}$ have degree 1, and 
the generators $\bar p_{\bar\mu}$ have degree 0. The extra generators can be understood 
as conjugate to the original generators of $\CA$. In the  Hochschild complex of multilinear 
maps, they correspond to the differential operators $\frac{\del}{\del z^\mu}$, 
$\frac{\del}{\del\bar z^{\bar\mu}}$ and $\frac{\del}{\del\bar\eta^{\bar\mu}}$, respectively. 
The conjugate relation can also be stated in terms of the Gerstenhaber structure of 
the Hochschild cohomology. This is endowed with an odd Poisson bracket, the Gerstenhaber bracket, 
which is given by the bidifferential operator 
\begin{equation}\label{brack}
\frac{\del}{\del z^\mu}\wedge\frac{\del}{\del\chi_\mu} + 
\frac{\del}{\del\bar z^{\bar\mu}}\wedge\frac{\del}{\del\bar\chi_{\bar\mu}} + 
\frac{\del}{\del\bar\eta^{\bar\mu}}\wedge\frac{\del}{\del\bar p_{\bar\mu}}. 
\end{equation}

When we take into account the differential $Q$ on the algebra $\CA$, we get an 
induced differential on the algebra $\CC$, which we called 
$\delta_Q$ before. It is given on the above polynomial algebra by 
\begin{equation}\label{deltaQ}
\delta_Q = \bar\eta^{\bar\mu}\frac{\del}{\del\bar z^{\bar\mu}}+\bar\chi_{\bar\mu}\frac{\del}{\del\bar p_{\bar\mu}} 
  \equiv \delta_1+\delta_2.
\end{equation}
To calculate the total Hochschild cohomology $H^*(\Hoch(\CA))$ we need to take the 
cohomology with respect to this differential.\footnote{A more precise procedure would 
involve a spectral sequence calculation for the double complex $\Hoch(\CA)$ with the 
total differential $\delta_m+\delta_Q$, for which the sketched procedure gives the second term.} 
Corresponding to the two factors, we write the above algebra as a tensor product 
\begin{equation}\label{split}
\CC = \C[z^\mu,\bar z^{\bar\mu},\bar\eta^{\bar\mu},\chi_\mu]\otimes
\C[\bar\chi_{\bar\mu},\bar p_{\bar\mu}].
\end{equation}
The idea is that the second factor is always trivial. The reason is that the variables 
$\bar\chi_{\bar\mu},\bar p_{\bar\mu}$ always generate a vector space, as they are coordinates 
on the fiber of the twisted holomorphic cotangent space to the 
$\bar z^{\bar\mu},\bar\eta_{\bar\mu}$-space. 
This should be contrasted with the first factor, where the generators $z^\mu,\bar z^{\bar\mu}$ 
should be interpreted as local coordinates on a topologically nontrivial manifold. 
The above polynomial algebra is only a local description of the full algebra $\CC$. 
As the two terms in the coboundary act on each factor separately, 
the cohomology is given by the tensor product of the cohomologies of the two factors. 
The second factor is easily seen to be completely 
trivial, and is given by\footnote{This can be realized by viewing it as the 
complexification of the cohomology of $\R^3$, parametrized by $\bar p_{\bar\mu}$, 
and $H^*(\R^3)=\R$ (supported at degree zero).}
\begin{equation}
H^*_{\delta_2}(\C[\bar\chi_{\bar\mu},\bar p_{\bar\mu}]) = \C.
\end{equation}
So we loose the second factor in \eqref{split}, and keep only the first factor. 
Also, we still have to take cohomology with respect to $\delta_1$. The 
total Hochschild cohomology therefore becomes 
\begin{equation}
H^*_{\delta_Q}(\CC) = H^*_{\delta_1}(\C[z^\mu,\bar z^{\bar\mu},\bar\eta^{\bar\mu},\chi_\mu]). 
\end{equation}
An argument similar to the one above would have us project out the 
$\bar z^{\bar\mu}$ and $\bar\eta^{\bar\mu}$ dependence. This would therefore 
not reproduce the full on-shell closed string algebra. But we should realize that the 
polynomial algebra only gives a local picture, in a contactable coordinate chart. 
And indeed the the cohomology in a local chart is expected to be trivial. 
Globally, one should find nontrivial cohomology. 

We can give a more precise global description as follows. The bosonic part is 
generated by the coordinates $z^\mu,\bar z^{\bar\mu},\bar p_{\bar\mu}$, which are 
coordinates on the total space of $\ol\CT_M^*$. Let us denote the projection 
on $M$ by $\pi:\ol\CT_M^*\to M$. 
The algebra $\CC$ can be described as sections of complex polyvector fields 
on the total space of $\ol\CT_M^*$, 
\begin{equation}
  \CC = \Gamma(\ext T(\ol\CT_M^*)\otimes\C) 
   = \Gamma(\ext\pi^*\CT_M\otimes\ext\pi^*\ol\CT_M\otimes\ext\pi^*\ol\CT_M^*). 
\end{equation}
Here we used that the vertical tangent space to $\ol\CT_M^*$ can be 
identified with $\pi^*\ol\CT_M^*$. 
This space can now be written as a tensor product of algebras over the 
algebra $\CO_M$ of complex functions on $M$,  
\begin{equation}
  \Gamma(\ext\CT_M\otimes\ext\ol\CT_M)
   \otimes_{\CO_M}\Gamma(\ext\pi^*\ol\CT_M^*). 
\end{equation}
The second part of the coboundary, $\delta_2$, acts only on the second 
part of this tensor product and commutes with the action of $\CO_M$. 
The correct way to calculate the total cohomology with respect to 
$\delta_Q=\delta_1+\delta_2$ would be to view the above again as a 
double complex, with the two terms as the two differentials, which we 
can calculate using spectral sequence techniques. The first term of this 
spectral sequence is the cohomology with respect to $\delta_2$, and 
as $H^*_{\delta_1}(\ext\pi^*\ol\CT_M^*)=\CO_M$ this is simply the 
first factor in the tensor product above, 
\begin{equation}
  E_1^{*,*}=\Gamma(\ext\CT_M\otimes\ext\ol\CT_M) = \Omega(M,\ext\CT_M). 
\end{equation}
The remaining coboundary is $\delta_1$, which is identified with 
the (twisted) Dolbeault operator $\bar\del$ acting on this space. 
Therefore, the second term in the spectral sequence now gives the 
required cohomology 
\begin{equation}
H^*(\Hoch(\CA)) = H^*_{\bar\del}(M,\ext[*]\CT_M),
\end{equation}
precisely the space of on-shell closed string operators. The sequence 
terminates, as we can always choose representatives that are independent 
of $\bar\chi$ and $\bar p$. 

We should note that the cohomology with respect to $\delta_m$ can be performed 
reliably in a local chart, due to the fact that the product is local. 
It is also important to note that indeed the cohomology of the second factor 
decouples globally. This is due to the fact that it involves the fibers only, 
which are globally trivial. A more precise argument would give the cohomology 
as the first term of a spectral sequence for the double complex formed by the 
terms $\delta_2$ and $\delta_1$. Because the $\delta_2$ cohomology is trivial, 
there is no room for descent equations, so we need not go further than the 
second term. 

If we consider the situation with a nontrivial gauge bundle $E$ 
and holomorphic connection $\bdA$. In fact, the algebra with values 
in $\End(E)$ is Morita equivalent to the algebra with trivial bundle. 
Furthermore, it is a well known result that the Hochschild cohomology 
is invariant under Morita equivalence. Therefore, the result does not 
depend on the gauge bundle.

\subsection{Identification of Cohomologies and Formality}

We see that the Hochschild cohomology $H^*(\Hoch(\CA))$ is precisely given 
by the BRST cohomology of the closed string theory, namely they can both 
be identified with the algebra $H^*_{\bar\del}(M,\ext[*]\CT_M)$. This 
gives confidence to our conjecture that in general the Hochschild cohomology 
calculates the on-shell closed string. 

This fact is actually important for the formality conjecture, in relation to 
path integral representations of this formality map, as in \cite{cafe,homa}. 
There it was implicitly assumed for formality that the two cohomologies are the 
same. Notice that formality is a statement about the relation between the 
Hochschild complex and its cohomology (namely, that they are quasi-isomorphic 
as $L_\infty$ algebras). On the other hand, the path integral gives a 
representation of a map from the closed string BRST cohomology to the Hochschild 
cohomology (an action on the $A_\infty$ algebra of the boundary theory). 
This can only be interpreted as formality if the cohomologies are the same. 

Formality of a complex (as a particular type of algebra) means that it is 
quasi-isomorphic to its cohomology.This means that there should be an 
intertwining map between the cohomology and the complex (or the other way around) 
that becomes an isomorphism in cohomology. 

More concretely, we have constructed a map from the closed string BRST 
cohomology to the Hochschild complex $\Hoch(\CA)$, as $\phi\mapsto\Phi$. 
This map is intertwining as shown in \cite{homa}. With the above calculation 
of the Hochschild cohomology we have shown that this map reduces to an isomorphism 
on the cohomology, as Gerstenhaber algebras algebras. In mathematical terms 
this means that the closed string BRST cohomology is quasi-isomorphic to the 
Hochschild complex. As the former can be identified with the cohomology of the 
latter, this reduces precisely to the mathematical notion of formality. 

This map $\phi\mapsto \Phi$ is only the first order approximation, 
in the context of the Hochschild cohomology. This can be seen by the fact 
that it is intertwining only to lowest order. Under certain conditions it 
can be extended to the full formality map $\phi\mapsto\bar\Phi$ which 
contains all higher order correction in the closed string field $\phi$. 
The components $\bar\Phi^{(n)}$ can be calculated similarly from the sigma 
model as the $n+1$-point functions completely deformed by $\phi$. 
These maps satisfy the full nonlinear master equation \eqref{master}. 

In general, there is an obstruction to extend the first order solution 
$\Phi$ in the Hochschild cohomology to a full solution $\bar\Phi$ of 
the master equation. Formality of the Hochschild complex says that 
there is a quasi isomorphism between the Hochschild cohomology, 
which has been shown to be the closed string BRST cohomology, 
and the Hochschild complex. This is a quasi isomorphism of 
$L_\infty$ algebras (and more generally, of $G_\infty$ algebras 
\cite{tam}). It therefore maps solutions of the master equation 
in the cohomology to solutions in the complex. As $\delta=\delta_Q+\delta_m$ 
is zero in the cohomology, the master equation there reduces 
simply to $\{\phi,\phi\}=0$. Hence this is a necessary condition 
for a full solution $\bar\Phi$ to exist. 

In the B-model, the master equation can be written in terms of 
the forms as 
\begin{equation}
  \frac{\del\phi}{\del z^\mu}\frac{\del\phi}{\del\chi_\mu} = 0. 
\end{equation}
This is nothing but the part that remains from the Gerstenhaber bracket 
\eqref{brack}. 
For a deformation of the complex structure $\varphi\in H^{-1,1}(M)$ 
this becomes 
\begin{equation}
  \del_\nu\varphi^\mu_{[\bar\mu}\varphi^\nu_{\bar\nu]} = 0.
\end{equation}
This can be understood as the quadratic part of 
$(\bar\del+\varphi\cdot\del)^2=0$. 
For an element $\theta\in H^{-2,0}(M)$ the master equation can be written 
\begin{equation}
  \theta^{\rho[\lambda}\del_\rho\theta^{\mu\nu]} = 0.
\end{equation}
This equation says that $\theta$ is a holomorphic Poisson structure. 
It is the condition for $\star_\theta$ to be associative to lowest order.

\section{A BV Sigma-Model}
\label{sec:bvmodel}

In this section we present a BV sigma-model giving an off-shell formulation 
for the B-model. This model is inspired by the calculation of the Hochschild 
cohomology in the previous section.

\subsection{The BV Model}

Above, we found the Hochschild cohomology of the open string B-model. In the intermediate 
step we found the algebra $\CC$ which was still provided with a nontrivial differential 
$\delta_Q$. Also, we saw that this algebra had a natural Gerstenhaber structure. Actually, 
this Gerstenhaber structure is easily seen to be part of a BV structure. That is, the 
bracket can be derived from the BV operator 
\begin{equation}
  \triangle = \frac{\del^2}{\del z^\mu\del\chi_\mu} + 
  \frac{\del^2}{\del\bar z^{\bar\mu}\del\bar\chi_{\bar\mu}} + 
  \frac{\del^2}{\del\bar\eta^{\bar\mu}\del\bar p_{\bar\mu}},
\end{equation}
as the failure of derivation condition. 

This BV structure can be used to define a 2-dimensional BV sigma model. To this end we 
introduce supercoordinates $(x^\alpha|\theta^\alpha)$ on the super worldsheet $\Pi T\Sigma$, 
where the $x^\alpha$ are the bosonic degree 0 coordinates on $\Sigma$ and the $\theta^\alpha$ 
are fermionic degree 1 Grassmann coordinates on the fiber. The superfields of the sigma model 
are superfields --- functions of the supercoordinates above --- corresponding to the 
generators of $\CC$. We will denote superfields using bold characters; 
they are given by 
\begin{equation}
\renewcommand{\arraystretch}{1.5}
\begin{array}{r@{\;}c@{\;}l}
  \Bz^\mu &=& z^\mu + \theta \rho^\mu + \theta^2f^\mu, \\ 
  \bar\Bz^{\bar\mu} &=& \bar z^{\bar\mu} + \theta\bar\rho^{\bar\mu} + \theta^2\bar f^{\bar\mu}, \\
  \bar\Bp_{\bar\mu} &=& \bar p_{\bar\mu} + \theta\bar\xi_{\bar\mu} + \theta^2\bar r_{\bar\mu}.
\end{array}
\qquad
\begin{array}{r@{\;}c@{\;}l}
  \Bch_\mu &=& \chi_\mu +\theta q_\mu + \theta^2\zeta_\mu, \\
  \bar\Bch_{\bar\mu} &=& \bar\chi_{\bar\mu} +\theta\bar q_{\bar\mu} + \theta^2\bar\zeta_{\bar\mu}, \\
  \bar\Bet^{\bar\mu} &=& \bar\eta^{\bar\mu} +\theta\bar h^{\bar\mu} + \theta^2\bar\pi^{\bar\mu}. 
\end{array}
\end{equation}
Here we did not explicitly write worldsheet form and vector indices and their contractions. 
On the space of superfields we have a BV structure induced by the BV operator 
$\triangle$, and a corresponding BV antibracket $(\cdot,\cdot)_{BV}$. 
The BV action of the sigma-model will be given by 
\begin{equation}
  S=\frac{1}{\kappa}\dint[_{\Pi T\Sigma}]{d^2\!xd^2\theta}\Bigl( \Bch_\mu D\Bz^\mu + \bar\Bch_{\bar\mu} D\bar\Bz^{\bar\mu} 
   + \bar\Bp_{\bar\mu} D\bar\Bet^{\bar\mu} + \bar\Bch_{\bar\mu}\bar\Bet^{\bar\mu} \Bigr),
\end{equation}
where $D=\theta^\alpha\del_\alpha$. 
The kinetic term in fact expresses the canonical relations between the generators. 
The last term induces the nontrivial differential on the $\CC$. In the BV language, 
this is the BRST operator, which is given by 
\begin{equation}
  \BQ = (S,\cdot)_{BV} = D+ \bar\Bet^{\bar\mu}\frac{\del}{\del\bar\Bz^{\bar\mu}} + \bar\Bch_{\bar\mu}\frac{\del}{\del\bar\Bp_{\bar\mu}}.
\end{equation}
This should be compared to the form of $\delta_Q$ \eqref{deltaQ}. In fact the 
potential term in the action was chosen precisely to reproduce this form. 
The last part of the structure of the BV sigma model is the 1-form operator 
$\BG_\alpha=\frac{\del}{\del\theta^\alpha}$. It corresponds to the operator 
$G_\alpha$, and satisfies the analogous relation $\{\BQ,\BG\}=d$. 

On the open worldsheet, the fields $\Bz^\mu,\bar\Bz^{\bar\mu},\bar\Bet^{\bar\mu}$ will satisfy 
Neumann like boundary conditions, while the fields $\Bch_\mu,\bar\Bch_{\bar\mu},\bar\Bp_{\bar\mu}$ 
satisfy Dirichlet boundary conditions. This implies that boundary operators 
are generated by functions of the first set of coordinates. These are precisely 
the observables of the open B-model. The observables in the bulk will be generated 
by functions of all superfields. These are pull backs to function space 
of the algebra $\CC$. The BRST cohomology of bulk operators will be related to the 
cohomology of $\CC$, which as we have seen precisely are the operators of the closed B-model. 

The gauge fixing above shows that operators have all kind of curvature corrections. 
It would be interesting to see if these survive the correlators. This would be 
somehow strange, as the B-model is supposed to be independent of the metric. 

This BV sigma model discussed in this section is closely related to similar 
BV models for the B-models given in \cite{alkoschza,js}. 
In fact they can be shown to be equivalent after partial gauge fixings.

\subsection{Gauge Fixing}

The BV sigma model defined this way can be seen to be equivalent to the 
usual B-model in BRST quantization, given by \eqref{baction}. 
To see this, we need to gauge fix the BV sigma model. To gauge fix we first 
need to make a division of the BV fields into ``fields'' and ``antifields''. 
We choose for the fields all of the scalars and the one-forms 
$\rho^\mu,\,\bar q_{\bar\mu},\,\bar h^{\bar\mu}$. To gauge fix the antifields in 
terms of the field, we use the following gauge fixing fermion, 
\begin{equation}
  \Psi = \int_\Sigma \Bigl(\kappa tg_{\mu\bar\mu}\rho^\mu*d\bar z^{\bar\mu} 
   -\frac{1}{2} \Gamma^\lambda_{\mu\nu}\rho^\mu\rho^\nu\chi_\lambda\Bigr). 
\end{equation}
This gives the gauge conditions 
\begin{equation}
\renewcommand{\arraystretch}{1.5}
\begin{array}{r@{\;}c@{\;}l@{\qquad}r@{\;}c@{\;}l}
 q_\mu &=& \kappa tg_{\mu\bar\mu}*d\bar z^{\bar\mu}
  - \Gamma^\lambda_{\mu\nu}\rho^\nu\chi_\lambda, &
 f^\mu &=& \displaystyle - \frac{1}{2}\Gamma^\mu_{\nu\lambda}\rho^\nu\rho^\lambda,\\
 \bar\zeta_{\bar\mu} &=& \displaystyle -\kappa tg_{\mu\bar\mu}d*\rho^\mu 
  - \frac{1}{2}R^\lambda_{\mu\bar\mu\nu}\rho^\mu\rho^\nu\chi_\lambda, &
 \zeta_\mu &=& \displaystyle \kappa t\del_\mu g_{\nu\bar\mu}\rho^\nu*d\bar z^{\bar\mu}
  - \frac{1}{2}\del_\mu\Gamma^\lambda_{\nu\tau}\rho^\nu\rho^\tau\chi_\lambda,
\end{array}
\end{equation}
with the other antifields vanishing. 
Inserting these defines the gauge fixed action. Taking the equations of 
motion for the auxiliary fields $\bar q_{\bar\mu}$ and $\bar p_{\bar\mu}$, 
we find $\bar h^{\bar\mu}\approx d\bar z^{\bar\mu}$. Substituting this 
in the gauge fixed action, we find back the original BRST action \eqref{baction}. 
Also, the operators $\BQ$ and $\BG$ reduce to the operators $Q$ and $G$ 
of the B-model respectively after gauge fixing. The covariantizing term 
in the gauge fixing term is similar to the one that can be included in the 
Cattaneo-Felder model \cite{balone}. 

Notice that the parameter $t$ originates from the gauge fixing. This gives 
another explanation why the theory does not depend on $t$, while $\kappa$ 
is a nontrivial coupling. 

The BV sigma model can also be used to do the calculations in this paper. 
Due to the natural symplectic structure the formulas are much more intuitive
The advantage of the BV model over the usual B-model is that the symmetry algebra 
generated by the BRST operator and $G$ closes off-shell. 
This is not true for the original model; for example we have 
$\{Q,G\}\chi_\mu=\kappa tg_{\mu\bar\mu}*d\bar\eta^{\bar\mu}$, which equals 
$d\chi_\mu$ only on-shell. Therefore, the BV sigma model can also be used 
to calculate off-shell amplitudes. This allows us to check that the expressions 
for the fundamental operations $\Phi^{(n)}_I$ remain correct for off-shell 
open string operators.

\subsection{Higher Orders and Kontsevich's Formula}

The BV action allows us to calculate correlators with more than one closed 
string insertions in a consistent perturbative expansion. Actually, our model 
is very close to the Cattaneo-Felder model \cite{cafe} used to calculate the 
star-product in deformation quantization. The perturbative expansion of 
their model essentially reproduced Kontsevich's formula for deformation 
quantization \cite{kon1}. The formula Kontsevich gave is a sum over graphs, 
which are identified with the Feynman graphs of the CF model, where each graph 
represents a particular term in the expansion, and has a particular calculable 
weight. The weight is essentially the value of the corresponding Feynman integral. 
In our case, adopting the gauge fixing procedure of Cattaneo-Felder (rather 
than the one above), we get essentially the same propagators and similar vertices. 
Hence the values of the Feynman integrals, and therefore the weights of the graphs, 
as in their model. The conclusion is that the perturbative expansion of the 
open-closed string field theory is essentially given by Kontsevich's formula. 
For example, this shows that indeed the nonlinear coupling of the $(-2,0)$ part 
$\theta$ of the closed string field combines to the deformed star-product 
$\star_\theta$, given by the (complexification of) Kontsevich's formula.

\section{Noncommutative Geometry of the B-Model}
\label{sec:ncg}

In this section we will interpret the results in terms of noncommutative geometry. 
It will be shown that the closed string algebra can be identified with 
with the cycles and chains in the sense of noncommutative geometry.

\subsection{Cyclic Cohomology}

Open string theories always have an inner product, which is defined by the 
2-point functions, which we assume to be nondegenerate. Therefore, we can relate 
multilinear maps with multilinear forms. We will denote the latter with 
$\Psi_I^{(n)}:\CA^{\otimes(n+1)}\to\C$ 
to distinguish them from $\Phi_I^{(n)}:\CA^{\otimes n}\to\CA$. They are defined by 
\begin{equation}
  \Psi_I^{(n)}(\hat\alpha_{a_0},\hat\alpha_{a_1}, \ldots, \hat\alpha_{a_n}) 
=  (-1)^{|\phi||\alpha_{a_0}|}\Vev{ \hat\alpha_{a_0} \Phi_I^{(n)}(\hat\alpha_{a_1}, \ldots, \hat\alpha_{a_n}) } 
= (-1)^{|\phi||\alpha_{a_0}|}\Phi^{(n)}_{Ia_0a_1 \ldots a_n}.
\end{equation}
Note that actually the forms are more natural than the maps, as they do not 
use the inner product. 

Open string correlators are in general cyclically antisymmetric, due to the fact that the 
boundary is closed. This will be true when  we insert the closed string operator 
$\phi_I$ provided we take $\del\phi_I=0$. 
In other words, the correlators $\Psi_I^{(n)}$ are (graded) cyclically 
antisymmetric, 
\begin{equation}
  \Psi_I^{(n)}(\hat\alpha_{a_0}, \ldots, \hat\alpha_{a_n}) 
  = (-1)^{n+\sum_{i>0}|\alpha_{a_0}||\alpha_{a_i}|}\Psi_I^{(n)}(\hat\alpha_{a_1}, \ldots, \hat\alpha_{a_n},\hat\alpha_{a_0}).
\end{equation}
One would therefore expect that the cohomology of these maps 
is the cyclic cohomology of the algebra $\CA$ rather than the Hochschild cohomology. 
The cyclic cohomology is a subcomplex of the Hochschild complex, consisting of 
the cyclically symmetric forms $\Psi$. 

For the trivial algebra $\CA=\C$, the cyclic cohomology can be found as follows. 
Let $e$ be the unit element generating the algebra $\CA=\C$. Then an element 
of cyclic cohomology is determined by the value $\Psi(e,\ldots,e)\in\C$. 
Cyclic symmetry restricts these to be zero for odd degrees $n$ (that is, an even 
number of arguments), therefore the cyclic cohomology is given by $\C$ in each 
even degree. In fact, it is generated as a polynomial algebra by a single element 
$\sigma$ of degree 2, $\HC^*(\C)\cong\C[\sigma]$. If $e$ is the unit element 
generating the algebra $\CA=\C$, the cyclic form is given by 
$\sigma(e,e,e)=1$.\footnote{This is exact in the Hochschild cohomology, as we can write 
$\sigma=\delta_m\tau$ where $\tau(e,e)=-1$. This immediately implies that it is closed 
also in the cyclic complex.} 
This is however not the closed string algebra, so indeed this is not the correct 
identification. The cyclic cohomology is however very close to the Hochschild cohomology. 
The relation can be given in terms of an exact triangle as follows \cite{connes}
\begin{equation}
\begin{array}{r@{}c@{}l}
& \HH^*(\CA) & \\ 
\raise5pt\hbox{\rlap{$\scriptstyle B$}}\swarrow && \nwarrow\raise5pt\hbox{\llap{$\scriptstyle I$}} \\ 
\HC^*(\CA) & \stackrel{S}{\relbar\joinrel\longrightarrow} & \HC^*(\CA)
\end{array}
\end{equation}
Here $I$ is the inclusion of the cyclic complex in the Hochschild complex. 
The map $B$ is the ``boundary map'', which is defined by 
\begin{equation}
  B\Psi(\alpha_{a_0},\ldots,\alpha_{a_n}) 
 = \frac{1}{n+1}\Bigl(\Psi(1,\alpha_{a_{0}},\ldots,\alpha_{a_{n}}) +(-1)^n\Psi(\alpha_{a_{0}},\ldots,\alpha_{a_{n}},1)\Bigr) \pm cycl.
\end{equation}
Notice that any (unital) algebra contains the algebra $\C$ as a subalgebra. Therefore, 
the element $\sigma$ naturally  is a canonical element of the cyclic cohomology 
of any algebra $\CA$. The map $S$ is the cup product with this element 
$S\Psi=\Psi\cup \sigma$. Note that $B$ lowers the degree by one and $S$ raises 
the degree by two. Therefore, in going around the triangle once, the degree 
is raised by one. The action of $S$ on the maps $\Phi$ can be written as follows 
\begin{eqnarray}
&&  \Vev{\hat\alpha_{a_0}\,(S\Phi_I)^{(n)}(\hat\alpha_{a_1},\cdots,\hat\alpha_{a_{n}})} \\
&&\qquad  = \sum_i\frac{\pm 1}{n+1}\Vev{\hat\phi_I\hat\alpha_{a_0}\int\!\hat\alpha_{a_1}^{(1)}\cdots
  \int\!\hat\alpha_{a_{i-1}}^{(1)}\hat\alpha_{a_{i}}\hat\alpha_{a_{i+1}}\int\!\hat\alpha_{a_{i+2}}^{(1)}\cdots
  \int\!\hat\alpha_{a_{n}}^{(1)}}
 \pm cycl.
\nonumber
\end{eqnarray}
Note that the therefore we included more general correlators. However, 
these correlators with less descendants will be factorizable. Therefore, 
they do not contain new information about the operators $\phi_I$. 
Furthermore it can be shown, compare \cite{connes}, that $S\Phi$ is 
trivial in the Hochschild cohomology. Therefore, it corresponds to a 
BRST exact closed string operator. 

The boundary map $B$ can be related with the BV operator in the closed 
string string theory. In fact, for the B-model this operator is identified 
with the $\del$ operator; that is it acts on the closed string 
operators $\phi$ we use as $(\triangle\phi)\cdot\Omega=\del(\phi\cdot\Omega)$. 
This can explicitly be seen from the expression we had in the calculation of the 
Hochschild cohomology. As the closed string operators can be taken independent 
of $\bar\chi$ and $\bar p$, the only term that survives is 
$\frac{\del^2}{\del\chi_\mu\del z^\mu}$, which is easily seen to exactly 
provide the above. Note that on the on-shell representatives we always can 
take $\triangle\phi=0$, and we actually assumed this. If we do not assume this, 
we indeed find that the explicit expression 
\begin{equation}
  \int\phi\cdot\Omega\wedge\Tr\Bigl(\alpha_{a_0}\wedge\del\alpha_{a_1}\wedge\cdots\wedge\del\alpha_{a_n}\Bigr)
\end{equation}
for the observables are not cyclically symmetric. Indeed we have to do a 
partial integration to move the $\del$ from $\alpha_{a_n}$ to $\alpha_{a_0}$. 
In general, we get an extra term involving precisely $\del(\phi\cdot\Omega)$. 
With the identification $B=(-1)^{p+q+1}\triangle$, we can write this as 
\begin{equation}
  \Vev{\alpha_{a_n}\alpha_{a_0},B\Phi(\alpha_{a_1},\cdots,\alpha_{a_{n-1}})} 
  = \pm \Vev{\alpha_{a_0},\Phi(\alpha_{a_1},\cdots,\alpha_{a_n})}
   \pm\Vev{\alpha_{a_n},\Phi(\alpha_{a_0},\cdots,\alpha_{a_{n-1}})}.
\end{equation}
Taking $\alpha_{a_n}=1$ this exactly reproduces the definition of the 
boundary operator $B$ above. It can also be checked that with the 
identification of $B$ as above also the signs are exactly reproduced. 

In the correlation functions, the reason that the correlation functions are 
not cyclic when $\triangle\phi\neq0$ comes from a correction in the Ward 
identity. Remember that the descent operator $G$ is generated by a 
current $\hat b$. The general action of this current on a closed string operator 
$\hat\phi$ is given by 
\begin{equation}
  \doint[_{C_z}]{\xi(w)\hat b(w)}\hat\phi(z) = \xi(z)\hat\phi^{(1)}(z)+\xi'(z)(\triangle\hat\phi)(z),
\end{equation}
where $\xi(z)$ is a vector on the worldsheet. In proving the Ward identity 
giving the cyclicity, we used a vector field $\xi$ vanishing at the point $z$ 
where the closed string operator was inserted. 

The relation between the cyclic cohomology and the Hochschild cohomology 
suggests the following interpretation in terms of the open string theory. 
The elements of the Hochschild cohomology correspond to the irreducible 
string diagrams, while the cyclic cohomology in general contains diagrams 
that can be factorized. The action on $\hat\phi$ will therefore vanish 
provided we take $\triangle\hat\phi=0$. If this is not the case, there 
will be a correction as $\xi'(z)$ can not be chosen to vanish. 
This correction will generate the extra term involving $B$. 

Next we remark on the structure of the observables. Note that we can write 
it in the form 
\begin{equation}
  \sum_{n\geq1}\frac{1}{(n+1)!}\int \Omega\cdot\Tr\Bigl(A(\triangle A)^n\phi\Bigr)+\cdots,
\end{equation}
where the dots denote the higher order corrections in $\phi$, and we 
should interpret $A$ as the full open string field including both 
the gauge field and the scalar. This expression can be interpreted as 
a potential for the deformation.

\subsection{Dirac Operator and Spectral Triples}

In our discussion on gauge invariant observables, we could think of the closed 
string operators $\phi$ as cycles over which we integrate the abstract 
1-form $A$. Actually, we can make this correspondence more exact using the language 
of noncommutative geometry \cite{connes}. In noncommutative geometry, cycles 
are elements of the cyclic cohomology. This is closely related to the Hochschild 
cohomology. We already saw above that the mixed correlators could be understood 
as elements of the cyclic cohomology. And therefore, the $\Phi_I$ can indeed be 
interpreted as cycles in the cyclic cohomology of the open string algebra $\CA$. 
In noncommutative geometry the cycles can be represented by characters, given by 
an integral $\int$ on a graded algebra $\Omega$ supplied with a map 
$\rho:\CA\to\Omega^0$ and a differential $d$, as 
\begin{equation}
  \tau(a_0,a_1,\cdots,a_n) = \int \rho(a_0) d\rho(a_1) \cdots d\rho(a_n).
\end{equation}
In the language of correlation functions, we can identify $d$ with $G$, 
$\Omega$ with the algebra generated by $\alpha_0\alpha_1^{(1)}\cdots\alpha_n^{(1)}$,
and the integral with the correlator with a $\phi$ insertion and an 
iterated integral over the boundary. The map $\rho$ is the obvious inclusion. 

More generally, Connes supplied a notion of metric on the space using 
a Dirac operator. What is needed is a spectral triple $(\CA,\CH,D)$, 
consisting of an algebra $\CA$, a Hilbert space $\CH$ and a Dirac operator $D$. 
The algebra $\CA$ and the Dirac operator $D$ act on the Hilbert space $\CH$, 
and there is the condition that $[D,a]$ is bounded for any $a\in\CA$ and 
$D$ has its spectrum in $\R$. 

We will now show that the B-model open string has a natural spectral triple. 
We already know that $\CA=\Omega^{0,*}(M,\End(E))$. Next we have to supply 
the Hilbert space $\CH$ and the Dirac operator $D$. As $\CA$ has to act on 
the Hilbert space, it is natural to take for $\CH$ the Hilbert space of 
$L^2$ sections of $E$-valued differential forms. At first, it seems that 
we could take $(0,q)$-forms. This is indeed enough for the differential 
geometry, but as we will see in order to have a Dirac operator we should take 
any $(p,q)$-form. A natural construction of the Dirac operator is to use 
a spinor bundle, and $D$ the usual Dirac operator on this spinor bundle. 
Let us therefore try to use the spinor bundle on $M$. As $M$ is a K\"ahler 
manifold (indeed, we assume it to be Calabi-Yau) a natural spinor bundle is 
provided by the bundle $\CS=\ext[*]\CT_M^*$ of $(p,0)$-forms. 
The Hilbert space and the Dirac operator are then given by 
\begin{equation}
 \CH = \Omega^{0,*}(M,\CS\otimes E)=\Omega^{*,*}(M,E),\qquad
 D = \del+\del^\dagger,
\end{equation}
where $\del$ and $\del^\dagger$ are the Dolbeault operators twisted by the 
bundle $E$, and the Hilbert space should be understood as a space of $L^2$ sections. 
The spinor bundle, and therefore also the Hilbert space $\CH$, has a 
natural $\Z_2$ grading induced by the degree. The Dirac operator 
is odd with respect to this degree. Note that indeed with an appropriate 
regularity for $\CA$ we have that $[D,\CA]$ is a subalgebra of the 
bounded operators on $\CH$. 

In noncommutative geometry, the Dirac operator can be used to recover the 
differential geometry. This is a particular representation of differential 
forms as 
\begin{equation}
  \alpha_0[D,\alpha_1]\cdots[D,\alpha_n]. 
\end{equation}
In fact one has to consider classes of them. Then cycles can be defined 
in terms of traces of these elements, 
\begin{equation}
  \Tr_\omega\Bigl(\alpha_0[D,\alpha_1]\cdots[D,\alpha_n]\Bigr). 
\end{equation}

Note that in the B-model we had a correspondence between operators 
in the string theory and the geometrical operations as 
\begin{equation}
  Q\sim\bar\del,\qquad Q'\sim\del^\dagger,\qquad 
  G_\parallel\sim\del, \qquad G_\perp\sim\bar\del^\dagger. 
\end{equation}
This came by identifying $\bar\eta^{\bar\mu}$ with the basic $(0,1)$-forms 
$d\bar z^{\bar\mu}$, and $\rho^\mu$ with the basic $(1,0)$-forms $dz^\mu$. 
We then see that the Dirac operator corresponds to the worldsheet operator 
$D \sim G_\parallel+Q'$.

\section{Other Models}
\label{sec:other}

We can do the same calculation for other topological string theories. 
We discuss here the trivial open string (with only the Chan-Paton degrees of freedom) 
the A-model and the Cattaneo-Felder model (C-model).

\subsection{The Trivial Model}

We start with the trivial model, whose only degrees of freedom are the Chan-Paton indices. 
The open string algebra is simply the matrix algebra $\CA=\Mat_N(\C)=\End(\C^N)$. 
This algebra is Morita equivalent to the algebra $\C$. As the Hochschild cohomology 
is invariant under Morita equivalence, this implies that $H^*(\Hoch(\CA))=H^*(\Hoch(\C))=\C$. 
This is indeed the on-shell closed string algebra, as the only operators of the 
closed string are multiples of the identity.

\subsection{The A-Model}

We next consider the topological A-model. The A-model is defined for a Lagrangian 
3-cycle in a Calabi-Yau manifold. Let us first assume that the Calabi-Yau is the total 
space of $T^*M$ for some real 3-manifold $M$ with $b_1(M)=0$, and the Lagrangian 3-cycle 
is the the base $M$. The open string 
field theory is now Chern-Simons on a real 3-manifold $M$. The open string algebra 
is given by $\CA=\Omega^*(M,\End (E))$ for some flat gauge bundle $E\to M$, with $Q=d_A$ 
the covariant derivative and the product is again the wedge-product. Locally for the 
trivial bundle, we can approximate the open string algebra as the polynomial algebra 
$\R[x^\mu,\eta^\mu]$, with $x^\mu$ the coordinates of $M$ and $\eta^\mu$ fermions of 
degree 1. As above, the Hochschild cohomology 
\begin{equation}
  \CC=H^*_{\delta_m}(\Hoch(\CA))
\end{equation}
can then be approximated by the polynomial algebra $\CC=\R[x^\mu,\eta^\mu;\chi_\mu,y_\mu]$, 
where $\chi_\mu$ have degree 1 and $y_\mu$ have degree 0. The coboundary operator is 
given by 
\begin{equation}
  \delta_Q = \eta^\mu\frac{\del}{\del x^\mu} + \chi_\mu\frac{\del}{\del y_\mu}.
\end{equation}
Again writing $\CC=\R[x^\mu,\eta^\mu]\otimes\R[\chi_\mu,y_\mu]$, and first taking 
cohomology with respect to the second term $\chi_\mu\frac{\del}{\del y_\mu}$ in the 
coboundary $\delta_Q$, the second factor has trivial cohomology. Therefore what 
remains is the first factor and the coboundary $\eta^\mu\frac{\del}{\del x^\mu}$, 
which is just the De Rham differential $d$ on forms. Globally, this gives the 
De Rham cohomology 
\begin{equation}
  H^*(\Hoch(\CA)) = H^*(M).
\end{equation}
We can use Morita equivalence to argue that the Hochschild cohomology does 
not depend on the choice of flat gauge bundle $E$. As the Calabi-Yau manifold 
is contractible to $M$, this is the same as the cohomology of the full 
Calabi-Yau $T^*M$, which is the closed string algebra.

In general this is however not precisely the closed string algebra. The latter is the 
cohomology of the Calabi-Yau space in which the 3-manifold $M$ is embedded 
as a Lagrangian cycle. We could have never found the full closed string, 
was the open string A-model as it stands knows only about the Lagrangian 
cycle $M$. There is a way to understand this, by remembering that the A-model 
is a topological subsector of the superstring on the Calabi-Yau. The Lagrangian 
submanifold $M$ is the space of a D-brane. This theory however has as its 
low-energy degrees of freedom not only the gauge field, but also 3 scalars, 
which represent the position of the D-brane in the Calabi-Yau. It is through these 
scalars that the decoupled wrapped D-brane knows about the bulk space. The solution 
therefore is to include in the topological open A-model also these 3 scalars. 
This can be done as follows. Note that the scalars $X$ together take values in the $NM$ 
normal bundle to $M$ in the Calabi-Yau. The K\"ahler form $\omega$ of the 
Calabi-Yau and the Lagrangian condition of the cycle $M$ defines 
an isomorphism of the normal bundle with the cotangent bundle of $M$, 
through $X\to\iota_X\omega$. We use this to define a complex gauge field 
$B=A+i\iota_X\omega$, in components $B_\mu=A_\mu+i\omega_{\mu i}X^i$. 
We propose a complex action which the usual 
Chern-Simons for this complexified gauge field. Note that because $X$ is 
in the adjoint of the gauge group, this still is invariant under the usual 
gauge transformations $B\to U\inv BU+iU\inv dU$. The complex action is actually 
invariant under the full complexified gauge symmetry, with the same formula for the 
gauge transformations but $U$ taking values in the complexified gauge group. 
The extra gauge transformations induce a nontrivial shift in the transverse 
coordinates, infinitesimally the new gauge transformations act on $X$ as 
$\delta_\Lambda X^i=[X^i,\Lambda]+\omega^{i\mu}\del_\mu\Lambda$.
Looking at the $\U(1)$ sector, and interpreting the scalars as 
embedding coordinates of the Lagrangian 3-brane, these shifts generate Hamiltonian 
flows with respect to the K\"ahler structure. These are indeed natural candidates for 
gauge transformations of this system. The complexified action can be interpreted 
as a superpotential, as in the case of the $B$-model. This also makes it plausible 
that the action becomes complex. The complex EOM found from this action are then 
interpreted as $F$-terms. Let us concentrate on the abelian model. 
The real part of this $F$-term is the usual condition for a flat connection 
$F=dA=0$. The imaginary part becomes $d(\iota_X\omega)=\CL_X\omega=0$.
This is actually equivalent to the condition that the 3-cycle shifted by the 
normal coordinates $X^i$ is still Lagrangian. 

To see that this can give the correct Hochschild cohomology we consider the case 
of a torus fibration over $M$. Let us assume that the total CY space looks like 
$T^*M$ with a nontrivial identification by a lattice $\Lambda\subset\R^3$ in the fiber, 
where we identified the fiber with $\R^3$. This implies that we have an 
identification of the scalars $X^i\to X^i+v^i$, for $v\in\Lambda$. 
As we understand $iX^i\omega_{i\mu}\eta^\mu$ as an element of the algebra $\CA$ 
it means that elements of this algebra should be identified modulo elements 
of the form $iv^i\omega_{i\mu}\eta^\mu$ for $v\in\Lambda$. For consistency, 
also products of such elements have to be identified with zero. 
After choosing coordinates such that $\omega\Lambda$ is identified with $\Z^3$ 
we can write these trivial elements as an integral subalgebra 
$i\Z[\eta^\mu]\subset\C[x^\mu,\eta^\mu]$. The identification of elements in $\CA$ 
modulo this subalgebra means that we should replace the open string algebra by the 
quotient algebra $\C[x^\mu,\eta^\mu]/i\Z[\eta^\mu]$. 

To see the effect of a quotient on the Hochschild cohomology, let us first consider 
the most trivial case where we have a single anticommuting generator $\eta$ and 
take the quotient algebra $\CA=\R[\eta]/\Z[\eta]$. We then approximate the Hochschild 
cohomology $\HH(\CA)$ using the HKR theorem introducing the dual variable 
$y\sim\frac{\del}{\del\eta}$. We still have to divide by the discrete algebra, so 
we have $\R[\eta,y]/\Z[\eta]$. We should require that the action 
of any element in the Hochschild complex acts trivially on the subalgebra $\Z[\eta]$ 
that we divided out. Taking the element $\e^{n\eta}\in\Z[\eta]$ for $n\in\Z$, we find 
\begin{equation}
  \phi(\eta,y)\e^{n\eta}=\e^{n\eta}\phi(\eta,y+1).
\end{equation}
This shows that the dual variable $y$ must be periodic. 
More generally, when we have several generators $\eta_\mu$ and consider a 
quotient algebra $\R[\eta_\mu]/\Lambda[\eta_\mu]$,  where $\Lambda$ 
is a lattice and $\Lambda[\eta_\mu]$ is the subalgebra degenerated by elements 
of the form $v^\mu\eta_\mu$ for $v\in\Lambda$, the Hochschild cohomology is 
described as the function algebra $\R[y^\mu,\eta_\mu]$ where $y^\mu$ are 
coordinates on a periodic plane with periods given by $\Lambda$, 
$y\sim y+v$. 

Going back to our original problem, we find that the dual generators $y_\mu$ 
in the Hochschild cohomology are periodic with according to the identification 
$y_\mu\to\omega_{\mu i} v^i$. This suggests that we should understand 
$x^i=\omega^{i\mu}y_\mu$ as the transverse coordinates on the total 
Calabi-Yau 3-fold. If we also understand $\eta^i=\omega^{i\mu}\chi_\mu$ as 
vertical 1-forms $dx^i$, we see that the Hochschild consists of all the 
(complex) forms on the full Calabi-Yau. Taking into account the BRST operator $Q$, 
we see that $\delta_Q$ is identified with the De Rham differential $d$ 
of the CY. This also shows that the total cohomology of the Hochschild 
double complex is precisely the complexified De Rham cohomology of the 
total space.

\subsection{The C-Model}

Similarly, we can do the same for the Cattaneo-Felder model \cite{cafe}, or C-model. 
The open string algebra is the algebra of functions $\CA=C(M)$ on a manifold $M$. 
The BRST operator is zero, and the product is the usual product of functions. 
Approximating functions by polynomials, we have $\CA=\C[x^\mu]$. Therefore the 
HKR theorem gives 
\begin{equation}
  \CC=H^*_{\delta_m}(\Hoch(\CA))=\C[x^\mu,\chi_\mu],
\end{equation}
where the generators $\chi_\mu$ have degree 1. This generates the algebra of 
polyvector fields $\Gamma(M,\ext[*]TM)$, by identifying the $\chi_\mu$ with the 
basis of vector fields $\del_\mu$. Because $Q=0$, this already equals the full 
Hochschild cohomology $H^*(\Hoch(\CA))$. Indeed, this is the on-shell closed 
string algebra of the C-model. 

The relation of this model to the deformation theory of the algebra 
of functions, through the Hochschild complex, was the main topic 
of the original paper \cite{cafe}. Given a bulk operator $\phi$ 
(called $\alpha$ in the paper) related to a polyvector field, they constructed 
polydifferential operators $U(\phi):C^\infty(M)^{\otimes n}\to C^\infty(M)$, 
defined through the path integral, 
\begin{equation}
  U(\phi)(f_1,\cdots,f_n)(x) = \Vev{\delta_x(X) (f_1(X)\cdots f_n(X))^{(n-2)}}_\phi,
\end{equation}
where the subscript $\phi$ indicates that the action is shifted by 
the term canonically related to $\phi$, and $\delta_x$ is the delta function 
at the point $x$. We see that the $U(\phi)$ can be identified with our 
$\Phi_\phi^{(n)}$ to first order in $\phi$.

\section{Conclusion and Discussion}
\label{sec:concl}

In this paper we have calculated mixed correlation functions in the topological 
B-model open-closed string theory. Our most important goal was to understand these 
in terms of deformations of the open string theory. These give the natural 
map from closed string BRST cohomology to the Hochschild cohomology 
of the open string algebra. By an explicit calculation of this 
Hochschild cohomology we found that in fact this map is an isomorphism. 
We conjectured in the introduction that this is true more generally, for 
any 2-dimensional topological field theory. In this paper we also 
checked this conjecture for the A-model in certain simple situations. 
One problem with this general conjecture is that it is not obvious what 
the closed string theory is given the open string sector. The conjecture 
could perhaps better be interpreted as a canonical construction of the latter. 

Although this calculation of the total cohomology of the Hochschild complex 
corresponds to the closed string BRST cohomology, we saw that the 
intermediate Hochschild cohomology $\HH^*(\CA)=H^*_{\delta_m}(\Hoch(\CA))$ 
contained all the closed string operators. More generally, the 
BV structure of this algebra allowed us to reproduce the full closed string 
B-model. Explicitly, we used the data to write down the BV sigma model 
for the B-model. 

In this paper we have considered mainly the B-model for a single D-brane. 
In general we can have bound states of several branes. These bound 
states can be considered as objects in the derived category of holomorphic 
sheaves $\CD(M)$ \cite{doug}. More generally, we can consider the full 
category of these boundary conditions at once. This catagory has an 
$A_\infty$ structure; though in a categorical sense. The Hochschild cohomology 
of $\CD(M)$ is in fact the Dolbeault cohomology, $H^{*,*}(M) \cong \HH^*(\CD(M),\CD(M))$. 
Also from this point of view the Frobenius structure of the closed string 
(the closed string action of \cite{bece}) is reproduced, see e.g. \cite{merk}. 
The fact that in general we should consider more objects at once also 
becomes clear when one looks at a cycle. In order to reproduce the full 
closed string, one needs at least a family of embedding curves. 
This became more important for the A-model, where all open string models 
are based on a cycle of lower dimension. This construction has an obvious 
relation, and indeed was inspired by, homological mirror symmetry 
\cite{kon5,bara1,bara2}. 

Another manifestation of this correspondence is related to the 
homology of loop space or string homology. It was shown \cite{cohjon} 
that it equals the Hochschild cohomology of 
the cochain complex $C^*(M)$, that is $H_*(LM) \cong \HH^*(C^*(M),C^*(M)^*)$. 
The loop space has an obvious relation with the closed string. 
This should actually be understood in terms of the C-model, 
where the full open string algebra indeed is the algebra of 
differential forms \cite{cafe}.

It would be very interesting to see if our conjectured relation 
between the Hochschild cohomology of the open string and the 
on-shell closed string algebra holds true in the case of the 
bosonic string. It is known that the bosonic open string field 
theory can be formulated in terms of a differential graded algebra. 
This can be very interesting in the context of tachyon condensation 
\cite{sen}. At this point, the open string field 
theory should describe the closed string vacuum, where only the closed strings are present. 
Note that this example shows the importance of using the full off-shell 
open string algebra, rather than its cohomology. Indeed, the cohomology 
in this case is trivial, while we expect a nontrivial Hochschild cohomology.
There are several problems however checking the proposal in this case. First of all, 
the (off-shell) open string algebra is very big, consisting of an infinite 
number of field components. Furthermore, the product structure, defined 
in terms of gluing half-strings, is much more complicated than the ones 
we saw in this paper. It can not be formulated in terms of the product in 
a polynomial or function algebra, therefore we can not use the HKR theorem. 

It was recently argued that in vacuum string field theory the (first order) 
coupling between the open and closed string is linear in the open string field 
\cite{grsz,hashitzh}. If our conjecture about the Hochschild cohomology holds 
true in VSFT this would imply that, at least with respect to a suitable basis, 
the Hochschild cohomology is completely concentrated in degree zero. 
Recently a much simpler expression for the star product was developed in 
\cite{dolimozw}. In this paper it was shown that, at least in the zero 
momentum sector, the open string star product reduces to a continuous 
product of Moyal products. This would reduce the calculation of the 
second term in the spectral sequence, the cohomology $\HH^*(\CA)$ 
with respect to $\delta_m$, to that of the ordinary star product on 
the noncommutative plane $\R^2_\theta$ for varying $\theta$. 
The latter cohomology is however trivial; it lives only in degree zero. 
This corresponds to operators which are linear in the open string field. 
This could therefore be an explanation why in open string field theory 
the coupling to the closed string should be linear.

\section*{Acknowledgments}

I am grateful to Whee Ky Ma for collaboration on related subjects 
and in initial stages of this paper. 
I like to thank Robbert Dijkgraaf, Erik Verlinde, Herman Verlinde, 
Jae-Suk Park, Yan Soibelman, Ezra Getzler, Hong Liu and Jeremy Michelson for useful discussions. 
I like to thank the IHES, where part of this work was conceived, for hospitality 
and support. This work was supported by DOE grant DE-FG02-96ER40959.

\appendix

\section{Gerstenhaber Structure of the Hochschild Cohomology}
\label{app:algebra}

\subsection{Gerstenhaber Algebras and BV Algebras}

A \emph{Gerstenhaber algebra} is a $\mathbb{Z}$-graded
algebra with a graded commutative associative product $\cdot$ of degree 0 and
a bracket $[\cdot,\cdot]$ of degree $-1$ (the Gerstenhaber bracket), 
which is such that $A[1]$ is a graded Lie algebra. This implies that 
it is twisted graded antisymmetric,
\begin{equation}
  [\alpha,\beta] = -(-1)^{(|\alpha|-1)(|\beta|-1)}[\beta,\alpha],
\end{equation}
and satisfies a twisted graded Jacobi identity, 
\begin{equation}
  [\alpha,[\beta,\gamma]] = [[\alpha,\beta],\gamma] +(-1)^{(|\alpha|-1)(|\beta|-1)}[\beta,[\alpha,\gamma]]
\end{equation}
Furthermore, the map $[\alpha,\cdot]$ must 
be a graded derivation of the product, 
\begin{equation}
  [\alpha,\beta\cdot\gamma] = [\alpha,\beta]\cdot\gamma
+ (-1)^{(|\alpha|-1)|\beta|}\beta\cdot [\alpha,\gamma]. 
\end{equation}
We can generalize this to a \emph{differential Gerstenhaber algebra} (or $DG$)
by adding a differential $\delta$ of degree 1, satisfying the graded derivation 
conditions with respect to the product and the bracket. 
We note that these identities are twisted graded variants of a \emph{Poisson algebra}. 

A \emph{BV algebra} is a Gerstenhaber algebra with a bracket of degree 1, 
(called the BV bracket) supplied with a nilpotent degree 1 operator $\triangle$ 
(the BV operator), $\triangle^2=0$, such that the bracket is given as the failure 
of its derivation property, 
\begin{equation}
  [\alpha,\beta] = (-1)^{|\alpha|}\triangle(\alpha\cdot\beta) 
  -(-1)^{|\alpha|}\triangle(\alpha)\cdot\beta - \alpha\cdot\triangle\beta.
\end{equation}

\subsection{Hochschild Complex and Cohomology}

The Hochschild complex of an associative algebra $A$ is defined in terms of the 
space of multilinear maps $C^n(A,A)=\Hom(A^{\otimes n},A)$. The coboundary 
of this complex is given in terms of the product $m=\cdot$ of $A$. 
For $\phi\in C^n(A,A)$ it is defined by 
\begin{eqnarray}
  \delta_m\phi(\alpha_0,\cdots,\alpha_n) &=& -\alpha_0\cdot\phi(\alpha_1,\cdots,\alpha_n)
  +(-1)^n\phi(\alpha_0,\cdots,\alpha_{n-1})\cdot\alpha_n \nonumber\\
&&  +\sum_{i=0}^{n}(-1)^{i}\phi(\alpha_0,\cdots,\alpha_{i}\cdot\alpha_{i+1},\cdots,\alpha_n).
\end{eqnarray}

It is well known that the Hochschild complex is endowed with a natural Gerstenhaber bracket. 
We first define a composition of two elements $\phi_i\in C^{n_i}(A,A)$ by 
\begin{equation}
 \phi_1\circ\phi_2(\alpha_1,\ldots,\alpha_{n_1+n_2-1}) = 
  \sum_i(-1)^{\epsilon_i}
 \phi_1(\alpha_1,\ldots,\alpha_i,\phi_2(\alpha_{i+1},\ldots,\alpha_{i+n_2}),\ldots,\alpha_{n_1+n_2-1}). 
\end{equation}
where $\epsilon_i=(n_2-1)i+\sum_{k=1}^i|\phi_2||\alpha_k|$. 
The Gerstenhaber bracket can now be defined as the graded commutator of this composition 
\begin{equation}
  [\phi_1,\phi_2] = \phi_1\circ\phi_2 -(-1)^{(n_1-1)(n_2-1)+|\phi_1||\phi_2|}\phi_2\circ\phi_1. 
\end{equation}
This can be interpreted as a double graded supercommutator, with the degrees 
$(n-1,|\phi|)$. These are indeed the natural gradings in the Hochschild complex. 
In additional there is a natural cup-product, defined by 
\begin{equation}
  \phi_1\cup\phi_2(\alpha_1,\cdots,\alpha_{n_1+n_2})
  = \phi_1(\alpha_1,\cdots,\alpha_{n_1})\cdot\phi_2(\alpha_{n_1+1},\cdots,\alpha_{n_1+n_2}).
\end{equation}
On cohomology, the cup-product can be shown to be symmetric. Together, the bracket
and the cup product it makes the Hochschild cohomology $\HH^*(A,A)=H^*_{\delta_m}(C^*(A,A))$ 
into a Gerstenhaber algebra. 

Note that if we interpret the product as an element $m\in C^2(A,A)$, we can write 
the action of the Hochschild coboundary on $\phi\in C^n(A,A)$ as 
$\delta_m\phi=(-1)^n[m,\phi]$. If $A$ would be a differential graded algebra, 
with differential $Q$ of degree 1, we could similarly introduce a second coboundary 
by the relation $\delta_Q\phi=[Q,\phi]$. This makes the Hochschild complex of a 
differential graded algebra into a double complex, with the natural bigrading 
mentioned above (such that $\delta_m$ has bidegree $(1,0)$ and $\delta_Q$ has 
bidegree $(0,1)$). More general, when there is a full $A_\infty$ structure 
with multiliner maps $m_k$, we have the differentials $\delta_{m_k}$ 
defined on $\phi\in C^n(A,A)$ by $\delta_{m_k}\phi=(-1)^{(k-1)n}[m_k,\phi]$. 
The $A_\infty$ relations are equivalent to the nilpotency of the total 
coboundary $\sum_k\delta_{m_k}$.

\subsection{The Hochschild Cohomology of a Polynomial Algebra}
\label{poly}

We can give an explicit description of the Hochschild cohomology of a
general polynomial algebra. Consider the algebra of polynomials in a
finite number of $\Z$-graded variables $x^i$ of degree
$|x^i|=q_i\in\Z$, so the space $\CA=\C[x^1,\ldots,x^N]$. 
The Hochschild cohomology of this algebra is,
as a $\Z$-graded vector space, the algebra of polynomials
$\HH^*(\CA)=\C[x^1,\ldots,x^N,y_1,\ldots,y_N]$ in the doubled set
of variables $x^i,y_i$, where the extra generators have degree
$|y_i|=1-q_i$ \cite{kon2}. The Gerstenhaber bracket of this 
polynomial algebra is given by $\frac{\del}{\del x^i}\wedge\frac{\del}{\del y_i}$. 
A representative of the cohomology class 
of any element of the cohomology in the Hochschild 
complex can be given by the polydifferential operator obtained by 
replacing $y_i$ with $\frac{\del}{\del x^i}$ (all differentiations 
acting on different arguments). For example 
$\theta(x,y)=\theta^{ij}(x)y_iy_j\in\HH^*(\CA)$ 
has as a representative the bidifferential operator 
$\theta^{ij}(x)\frac{\del}{\del x_i}\wedge\frac{\del}{\del x_j}$. 
This result is known as the Hochschild-Kostant-Rosenberg theorem. 
More generally, for the algebra $\CA=\CO(M)$ of
regular functions on a smooth $\Z$-graded algebraic supermanifold $M$,
the Hochschild cohomology is given by the algebra of functions on the
total space of the twisted by $[1]$ cotangent bundle to $M$,
$\HH^*(\CO(M))=\CO(T^*[1]M)$. Because the new space is a cotangent 
bundle, it has a natural Gerstenhaber bracket, where the degree 
of $-1$ is a result of the twisting. 

There is also an analytic version of this theorem, which is due to Connes.

\end{document}